\ifdefined\submissionlayout
  \documentclass[aip,pop,preprint,12pt,groupedaddress]{revtex4-1}
\else
  \documentclass[aip,pop,reprint,groupedaddress]{revtex4-1}
\fi
\usepackage{dcolumn}
\usepackage{amsmath}
\usepackage{bm}
\usepackage{hyperref}
\usepackage{natbib}
\usepackage{amsfonts}
\usepackage{booktabs}
\usepackage{siunitx}
\usepackage{relsize}
\usepackage{mathtools}
\usepackage{physics}
\usepackage{graphicx}
\usepackage{array}
\usepackage{enumitem}

\makeatletter
\def\namedlabel#1#2{\begingroup
    #2
    \def\@currentlabel{#2}
    \phantomsection\label{#1}\endgroup
}
\makeatother

\usepackage{color}
\usepackage{amssymb}

\newcommand{\D}{\textrm{D}}

\AtBeginDocument{\def\natexlab#1{}}

\usepackage{microtype}
\AtBeginDocument{\setcitestyle{numbers,square}}
\AtBeginDocument{}
\ifdefined\submissionlayout
  \let\origincludegraphics\includegraphics
  \renewcommand{\includegraphics}[2][]{{\setlength{\columnwidth}{8.5cm}\origincludegraphics[#1]{#2}}}
\fi

\newcommand{\ee}{\mathrm{e}}
\newcommand{\ii}{\mathrm{i}}
\newcommand{\nn}{\mathrm{n}}
\newcommand{\kB}{k_\mathrm{B}}
\newcommand{\mee}{m_\mathrm{e}}
\newcommand{\mi}{m_\mathrm{i}}
\newcommand{\nee}{n_\mathrm{e}}
\newcommand{\Te}{T_\mathrm{e}}
\newcommand{\Ti}{T_\mathrm{i}}
\newcommand{\Tn}{T_\mathrm{n}}
\newcommand{\ub}{\mathbf{u}}
\newcommand{\Jb}{\mathbf{J}}
\newcommand{\Eb}{\mathbf{E}}
\newcommand{\Bb}{\mathbf{B}}

\newcommand{\sP}{\sigma_\mathrm{P}}
\newcommand{\sH}{\sigma_\mathrm{H}}
\newcommand{\spar}{\sigma_\parallel}
\newcommand{\Kn}{\mathrm{Kn}}
\renewcommand{\dd}{\mathrm{d}}
\newcommand{\Ha}{\mathrm{Ha}}

\begin{document}

\title{A multifluid model of hydrogen isotope separation in a partially ionized plasma centrifuge}

\author{S. Cao}
\author{M. D. Acciarri}
\affiliation{Marathon Fusion, San Francisco, CA, USA\\ \textup{Corresponding author: S. Cao, \href{mailto:lennox@marathonfusion.com}{lennox@marathonfusion.com}}}

\begin{abstract}

We present an axisymmetric, time-dependent multifluid model of a
partially ionized plasma centrifuge and test it against the
hydrogen-deuterium separation experiments of Korobtsev \textit{et al.} The model reproduces the
measured separation factors to within 26\,\% at all five discharge currents (Fig.~\ref{fig:alpha}). Unlike in earlier models, the
properties of the plasma and the gas, such as the temperatures and ionization
fraction, are computed rather than imposed from measurements or adjusted as free
parameters. The main
uncertainties of the model are the rarefaction of the gas in the core, the neglect of ion
advection in the transport of the plasma density, and the omission of molecular
processes, which will be the focus of future work.
\end{abstract}


\maketitle

\section{Introduction}
\label{sec:intro}

In a plasma centrifuge a radial current crossed with an axial magnetic field
exerts an azimuthal force on charged species. In the partially ionized variant, the
ions pass this momentum to the neutral gas by charge-exchange and elastic
collisions. The currents then act like an immersed, weightless
rotor~\cite{Lehnert1973}, and the rotation is not limited by the strength of a
wall material. The concept dates from the 1960s. Bonnevier derived the radial
separation of ion species by ion-ion friction in a fully ionized rotating
plasma~\cite{Bonnevier1966} and later measured element and isotope separation
in a pulsed rotating discharge~\cite{Bonnevier1971}. Lehnert reviewed rotating
plasmas, including the critical ionization velocity that may cap the
plasma-neutral slip~\cite{Lehnert1971}, and estimated the operating conditions,
power and separation of the partially ionized
centrifuge~\cite{Lehnert1973}. Partially ionized devices were then built with
steady glow discharges in He-Ar~\cite{Kaneko1978}, with arcs in Ar-Xe at 1 to 3~Torr~\cite{Wijnakker1979,Wijnakker1980thesis},
and with steady arcs in
argon~\cite{Simpson1987}. The pulsed centrifuge of Korobtsev \textit{et al.}
separated H from D and hydrogen isotopes from neon at a few tenths of a
Torr~\cite{Korobtsev1984,korobtsev1986chemical}. In that device the rotation
speed, the gas temperature, the plasma density and the separation factor were
all measured, which makes it the benchmark for this work. Several of these
experiments reported rotation or separation well below simple theoretical
estimates~\cite{Kaneko1978,Simpson1987,SimpsonYue1999}, and the gas temperature was repeatedly identified as the limiting
quantity~\cite{Wijnakker1980,Wijnakker1980thesis,SimpsonYue1999}.

Mass-selective separation is of interest for the fuel cycle of
deuterium-tritium fusion. Whyte \textit{et al.}~\cite{Whyte2023} showed that
the tritium burn efficiency
is controlled by the helium fraction allowed in the divertor and by the ratio of
the effective pumping speeds of helium ash and of the unburned fuel, so that
pumping the ash faster than the fuel is effective. Parisi \textit{et
al.}~\cite{Parisi2024} proposed pumping deuterium and tritium asymmetrically
for the same purpose. A partially ionized centrifuge working on the atomic
exhaust gas is one candidate for such mass-selective pumping.

This
paper concerns the modeling of the centrifuge itself: we develop a coupled
model, state its strengths and limitations, and test it against established H/D
separation experiments. Mixtures of deuterium, tritium and helium are future work.

Previous models were not self-consistent and relied on imposed or measured
quantities as inputs. Bonnevier and
Lehnert worked analytically, with prescribed rotation and temperature profiles,
or with the rotation fixed at the critical
velocity~\cite{Bonnevier1966,Lehnert1971,Lehnert1973}. Wijnakker \textit{et
al.} solved a linear, homogeneous single-fluid problem for the azimuthal
velocity and the potential in the $(r,z)$ plane, with $r$ the radial and $z$ the axial coordinate, using measured temperatures and
densities and a Hall parameter fitted to the arc voltage~\cite{Wijnakker1979}.
They then solved a one-dimensional radial energy balance with a prescribed
rotation profile. With viscous dissipation as the only heat source this balance
caps the ratio of rotational to thermal energy, $X=\tfrac12mu_\theta^2/\kB T$ (particle mass $m$, azimuthal velocity $u_\theta$, temperature $T$, Boltzmann constant $\kB$), below unity, although Ohmic heating was
estimated to be at least six times larger~\cite{Wijnakker1980}. Adding Ohmic
and electron-collision heating with a prescribed electron temperature, Wijnakker
concluded that the electron energy equation had to be part of a more accurate
theory~\cite{Wijnakker1980thesis}. Van den Berg solved a problem of the same kind for the same device by finite
elements in the $(r,z)$ plane~\cite{VanDenBerg1982,Bakker1980}. He used a
uniform effective Hall parameter that includes ion slip and is computed from
collision frequencies, rather than fitted, over a range of pressures and fields.
The meridional flow and inertia were set to zero; the viscosity, the
conductivity, the temperatures and the plasma density were uniform; and the
energy equation was excluded. The computed rotation peaked near the cathodes
rather than at the midplane, contrary to the measurements, and the computed
anode-cathode voltage was considerably larger than the measured one, which he
attributed to the uniform temperature. The secondary circulation was obtained
only qualitatively, from a linearized vorticity equation driven by the
zero-order centrifugal pressure.
Simpson's three-fluid model resolved the Hartmann layer on the end walls at a
single radius, with an ambipolar particle balance. It needed a prescribed ionized
fraction and an anomalously enhanced ion-electron collision frequency to match
measured voltage-current curves~\cite{Simpson1981b}. Its successor, a
steady one-dimensional model of the end-wall boundary layers for a single argon
fluid in local thermodynamic equilibrium, computes the temperature from the
energy equation with a specified wall temperature; the core speed $v_0=-E_r/B$, with $E_r$ the radial electric field and $B$ the magnetic field, is a
constant chosen to match the current, and a 5\,\% non-uniformity of the
conductivity is assumed~\cite{Yue1996}. More recently, Wong \textit{et
al.}~\cite{Wong2024} followed the transient radial dynamics of a rotating weakly
ionized plasma with a charged fluid and a neutral fluid at a prescribed
temperature, without electrons or ionization, and Muir \textit{et
al.}~\cite{Muir2025} derived steady one-dimensional rotational profiles of a weakly
ionized hydrogen plasma with Saha ionization and a single prescribed temperature,
without an energy equation. Reviews of plasma isotope and mass separation, including the
competition of circulation and diffusion, are given in
Refs.~\cite{Zweben2018,Dolgolenko2023}.
None of the models named above contains more than one neutral
species as a separate fluid. The separation factor is either not computed, or is obtained from the
isothermal centrifugal equilibrium with a velocity and temperature common to all
species. Simpson's three-fluid model explicitly excludes
mixtures~\cite{Simpson1981a}, its successor treats a single fluid~\cite{Yue1996},
and the separation measured by Korobtsev
\textit{et al.} was interpreted with order-of-magnitude
estimates~\cite{Korobtsev1984}.

The attainable separation depends on $u_\theta^2/T$. The gas temperature
results from Joule heating of electrons and ions, collisional energy exchange,
viscous dissipation, conduction to the walls and convection by the secondary
flow, and all of these depend on the rotation and on the plasma. A model that
prescribes the temperature, the rotation or the plasma state therefore cannot
predict the separation factor from experimental parameters alone. The model presented here couples these
processes. It is time-dependent and axisymmetric in the $(r,z)$ plane, and no
measured rotation, gas temperature or ionization degree is imposed: the inputs
are the geometry, the fill gas, the magnetic field, the wall conditions, the
collision data, the seed plasma and the discharge current $I(t)$ as a function of the time $t$, with its
profile on the cathode. Its distinctive features are the combination of a
time-dependent two-dimensional discharge with a computed thermal evolution,
separate neutral isotopes and the circulation driven by the end walls, and the comparison
with pulsed H/D separation measurements. The electrostatic potential and the current
distribution follow from current continuity with the full anisotropic
conductivity tensor, evaluated from the local plasma state. The plasma
density and the electron temperature evolve with ionization, recombination and
the radiation that accompanies them, all taken from the collisional-radiative
rate coefficients of the AMJUEL database~\cite{ReiterAMJUEL}, so the
ionization degree and the conductivity are computed rather than assumed. The
ions have their own temperature and azimuthal momentum equation, which resolves
the ion-neutral slip. Every neutral species is treated as a separate compressible
Navier-Stokes fluid, coupled to the other species by binary friction and to the
plasma by drag and by the Lorentz force. The separation thus emerges from the
slip between species. It is not assumed to follow a local equilibrium, and it
includes transients and the secondary circulation. The Joule heating is
resolved by species in the frame of the gas, and the gas temperature is
computed.
Because the model is time-dependent, it can be compared directly with pulsed
experiments. It reproduces the separation factors measured by Korobtsev
\textit{et al.}~\cite{korobtsev1986chemical} to within 26\,\% at all five
discharge currents (Fig.~\ref{fig:alpha}).

The fluid treatment also has limits. First, the neutral mixture is described as a continuum,
although the heated core of the simulated discharges is rarefied. Second, quasineutrality is
imposed everywhere, including the electrode layers. Third, the plasma density is transported by
ambipolar diffusion only, without advection by the ion drift. Fourth, molecules are not included.
Section~\ref{sec:validity} discusses these approximations and their consequences.

The paper is organized as follows. Section~\ref{sec:device} describes the device, the
experiments and the simulated configuration. Section~\ref{sec:model} gives the
governing equations and Section~\ref{sec:numerics} the numerical method and its
verification. Section~\ref{sec:scaling} collects the scaling estimates against which
the simulations are read. Section~\ref{sec:results} presents the results, Section~\ref{sec:validity} discusses the limits of validity of the fluid description,
and Section~\ref{sec:conclusions} concludes. The supplementary material contains the
verification of the gas solver.

\section{Device, experiments and simulated configuration}
\label{sec:device}

\begin{figure*}[t]
\centering
\includegraphics[width=0.8\textwidth]{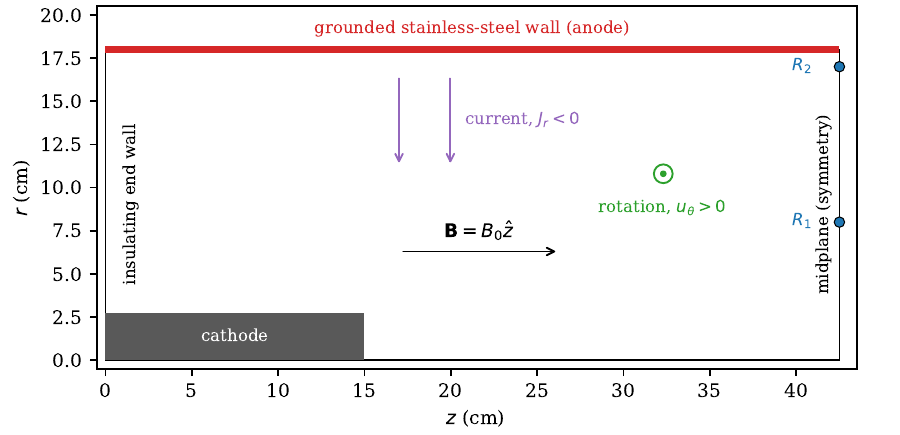}
\caption{Computational domain of the Korobtsev device model, drawn to scale.
Half of the machine is simulated, by symmetry about the midplane. The dots mark
the sampling radii of the experiment on the midplane. The arrows show the magnetic
field and the radial current; the circled dot marks the rotation of the gas, which is out of the page.}
\label{fig:domain}
\end{figure*}
\subsection{Operating principle}

A cylindrical vessel is immersed in a uniform axial magnetic field
$\Bb=B_0\hat{\mathbf z}$ of strength $B_0$, where $\hat{\mathbf z}$ is the unit vector along the axis
of the cylindrical coordinates $(r,\theta,z)$, and a radial current flows between a central
electrode and the vessel wall (Fig.~\ref{fig:domain}). With the central electrode as cathode, the radial
current density is $J_r<0$: current flows inward and electrons outward. The
azimuthal force density $-J_rB_0>0$ acts on the charged species, which hand it
to the neutral gas by ion-neutral friction (mostly resonant charge exchange)
and electron-neutral friction.

With the current prescribed, the total Lorentz torque is fixed by the current
and the magnetic field, and does not depend on the plasma state. How that torque and the Joule power are shared between plasma and gas sets the
rotation, the gas temperature and density, and finally the separation.
The separation that a given rotation and temperature allow is estimated in Section~\ref{sec:scaling-alpha}.

\subsection{The experiments of Korobtsev et al.}
\label{sec:experiment}

The discharge chamber is a stainless-steel cylinder 36~cm in diameter and 85~cm
long~\cite{Korobtsev1984,korobtsev1986chemical}; the device is symmetric about its midplane, and Fig.~\ref{fig:domain} shows one half. Two central electrodes with
hemispherical tips, 5.5~cm in diameter, are introduced along the axis through
quartz insulators, 15~cm into the chamber from each end. A solenoid produces an
axial field of up to 0.5~T. The radial current is driven by a 1575~$\mu$F
capacitor bank. The discharge lasts 10 to 15~ms, with peak currents of 0.4 to 2~kA.
The polarity of the H/D measurements is not stated. We take the central electrode as the cathode and the vessel wall as the anode. We assume that each cathode receives the quoted current.

The separation measurements simulated here were made in a 50/50 H$_2$/D$_2$ mixture at a fill pressure $p_0=0.3$~Torr and a field of 0.5~T, for peak currents of 0.76 to 1.55~kA~\cite{korobtsev1986chemical}. The gas was sampled through probes at the radii $R_1=8$~cm and $R_2=17$~cm on the midplane (Fig.~\ref{fig:domain}), 2.5~ms after the start of the discharge.
The mean plasma density, measured by microwave interferometry and electric probes, was $(1$ to $2)\times10^{19}$~m$^{-3}$ at 1~kA; the region over which it was averaged is not stated. The rotation, measured from the Doppler shift of spectral lines and therefore an average along the line of sight, was about 13~km\,s$^{-1}$ in hydrogen at 1~kA and 0.5~T, at an unstated time.

\subsection{Simulated configuration}
\label{sec:config}

By symmetry, we simulate half of the device (Fig.~\ref{fig:domain}): $0\le r\le18$~cm and $0\le z\le42.5$~cm, with the end plate at $z=0$ and a
symmetry condition at $z=42.5$~cm. The electrode is represented by a flat-topped
cylinder of radius 2.75~cm protruding 15~cm into the chamber and acts as the
cathode; the stainless-steel wall is the grounded anode, and the end wall is
insulating. H$_2$/D$_2$ is represented by atomic H and D at twice the molecular
number density, which preserves the mass density. The discharge dissociates
the molecules almost completely~\cite{korobtsev1986chemical}, and molecular
processes are not included: there are no molecular ions, no dissociation, and
no recombination of H and D atoms on the walls into H$_2$, HD and D$_2$. Atoms
and ions that reach a wall return to the gas as atoms. The consequences of this
simplification are discussed in Section~\ref{sec:validity-molecules}. The measured current
waveform of the device is not published. We therefore use a candidate
waveform built from what the papers state: a critically damped discharge of
the 1575~$\mu$F bank,
\begin{equation}
I(t)=I_\mathrm{peak}\,\frac{t}{\tau}\,\exp\!\left(1-\frac{t}{\tau}\right),
\qquad\tau=3~\mathrm{ms},
\label{eq:pulse}
\end{equation}
which rises to its peak $I_\mathrm{peak}$ at the time $t=\tau$, 3~ms after initiation, the time of the current maximum given for the same device in Ref.~\cite{Korobtsev1984}. With the stated
capacitance $C$ it corresponds to a circuit inductance $L=\tau^2/C\approx5.7$~mH and a circuit resistance
$R=2\tau/C\approx3.8~\Omega$, and it falls to 20\,\% of its peak at 12~ms and
to 9\,\% at 15~ms, consistent with the stated duration of 10 to 15~ms. The peak
currents are 0.76, 0.96, 1.25, 1.37 and 1.55~kA, the five currents of the H$_2$-D$_2$ measurements, and the simulations stop at 3.5~ms.
The
rise of the current is not published, and the separation at 2.5~ms depends on
the current history before it, so this waveform is a source of uncertainty in
the comparison. Gas and walls are initially at 300~K, and the seed plasma
is uniform, with an electron density $\nee=10^{18}$~m$^{-3}$ and an electron temperature $\Te=0.25$~eV.

The simulated plasma is in a low-temperature, partially ionized regime with strongly magnetized electrons. On the midplane at the current peak, for the five currents, 1 to 4\,\% of the atoms are ionized; averaged over the midplane, the electrons and the gas are both at about 1~eV, and the electron Hall parameter is 6 to 27. The ion Hall parameter varies across the midplane, from
about 0.1 near the wall to 0.6 to 0.9 at $r=8$~cm and about 3 at 4~cm, in the rarefied
core.
On the midplane the ions carry about 40 to 80\,\% of the cross-field current.

\section{Model}
\label{sec:model}

The domain is axisymmetric, $\partial_\theta=0$. Table~\ref{tab:unknowns} lists the unknowns of the model.

\begin{table}[t]
\centering
\caption{The unknowns of the model and the equation that determines each. The subscript $\nn$
denotes the neutral gas as a mixture of the neutral species $s$.}
\label{tab:unknowns}
\setlength{\tabcolsep}{4pt}
\begin{tabular}{lll}
\hline\hline
Symbol & Definition & Equation \\
\hline
\multicolumn{3}{l}{\textit{Unknowns}}\\
$\phi$ & electrostatic potential & \eqref{eq:phi} \\
$\nee$, $\Te$ & electron density and & \eqref{eq:ne}, \eqref{eq:Te} \\
 & temperature & \\
$\Ti$, $u_{\ii\theta}$ & temperature and azimuthal & \eqref{eq:Ti}, \eqref{eq:uitheta} \\
 & velocity of a single singly & \\
 & charged ion fluid & \\
$\rho_s$, $\ub_s$, $T_s$ & density, velocity and & \eqref{eq:neutral_mass} to \eqref{eq:neutral_E} \\
 & temperature of one fluid per & \\
 & neutral species $s$ & \\
\multicolumn{3}{l}{\textit{Neutral gas as a mixture}}\\
$n_\nn$, $\ub_\nn$, $\Tn$ & total number density, & \\
 & mass-averaged velocity and & \\
 & number-averaged temperature & \\
\hline\hline
\end{tabular}
\end{table}

\subsection{Assumptions}
\label{sec:assumptions}

The model rests on the following approximations:
\begin{enumerate}
\item \textbf{Forced quasineutrality.} The ion density equals the electron density, $n_\ii=\nee$, everywhere. Sheaths are
  not resolved and enter only through boundary conditions. Electrode layers
  that in reality are thinner than a cell of the computational grid (Section~\ref{sec:numerics}), or non-neutral, are represented by
  the first fluid cell.
\item \textbf{Electrostatic field, uniform $\Bb$.} The magnetic field is
  $\Bb=B_0\hat{\mathbf z}$ everywhere. In the simulations the diamagnetic
  change of $B_z$ stays below 2\,\% of $B_0$, and the self field $B_\theta$ below
  5\,\%, the maximum being just above the cathode tip. The papers describe the field of
  the solenoid as uniform and give no spatial profile, so a uniform field is
  assumed.
\item \textbf{Drift closure for the charged species in the $(r,z)$ plane.}
  The electron and ion $(r,z)$ velocities follow from the local drag balance,
  that is, from Ohm's law (Section~\ref{sec:ohm}). Only the ion azimuthal velocity
  keeps inertia, friction and viscosity.
\item \textbf{Ambipolar diffusion without ion advection.} The plasma density
  obeys a diffusion equation with the ambipolar coefficients
  (Section~\ref{sec:plasma}).
\item \textbf{Maxwellian electrons.} The rate coefficients are Maxwellian averages at
  $\Te$; the tail of the distribution that sets ionization and excitation is assumed
  thermal (Section~\ref{sec:plasma}).
\item \textbf{Continuum neutrals.} Each neutral species obeys the compressible
  Navier-Stokes-Fourier equations with Chapman-Enskog transport
  (Section~\ref{sec:neutral}). The validity of this assumption is discussed in
  Section~\ref{sec:validity-neutral}.
\end{enumerate}

\subsection{Ohm's law and current continuity}
\label{sec:ohm}

In the simulated discharges the ionization fraction ranges from less than 1\,\% in the outer gas to more than 50\,\% in the core, so electron-ion collisions are negligible in one part of the current path and dominant in another. Many of the conductivities in common use for weakly ionized plasmas neglect these collisions, so they do not apply in the core, where electron-ion collisions dominate. There, the friction between the electrons and the ions couples their momentum balances, which then have to be solved together, as Song \textit{et al.}~\cite{Song2001} showed for the ionosphere. In the frame of the gas, however, they give the conductivities only in an approximate form. We therefore derive them here, solving the two balances exactly in that frame.

The current follows from the momentum balances of the electrons and the
ions in the drift limit. With their inertia and viscous stress dropped, each
charged species $j$ balances the Lorentz force against the friction on the other
charged species $j'$ and on the gas,
\begin{multline}
0=Z_j\ee\nee(\Eb+\ub_j\times\Bb)\\
-m_j\nee\big[\nu_{jj'}(\ub_j-\ub_{j'})+\nu_{j\nn}(\ub_j-\ub_\nn)\big],
\label{eq:mom}
\end{multline}
where $\ee$ is the elementary charge, $\Eb$ is the electric field, $Z_j$ is the charge number, $Z_\ii=1$ and $Z_\ee=-1$, $\ub_j$ and $m_j$ are the velocity and the particle mass of species $j$,
and $\nu_{jj'}$ and $\nu_{j\nn}$ are the frequencies of its momentum-transfer collisions with the other charged species and with the gas
(Section~\ref{sec:collisions}). The mass of the ion fluid is the mean
atomic mass of the local gas,
\begin{equation}
\mi=\frac{n_\mathrm{H}m_\mathrm{H}+n_\D m_\mathrm{D}}{n_\mathrm{H}+n_\D}
\label{eq:mi}
\end{equation}
for the H/D fill, where $n_\mathrm{H}$ and $n_\D$ are the number densities and $m_\mathrm{H}$ and $m_\mathrm{D}$ the atomic masses of the two isotopes, because resonant charge exchange keeps the isotope ratio of
the ions at that of the gas around them. We leave out the pressure force $-\nabla p_j$:
without ion advection in the plasma density (Section~\ref{sec:plasma}), the
pressures of the electrons and the ions are not accurate enough to include
it. The friction between electrons and ions acts on
their relative drift and exerts no force on the gas; momentum conservation
gives $\mi\nu_{\ii\ee}=\mee\nu_{\ee\ii}$.

We write the drifts relative to the gas, $\mathbf w_j=\ub_j-\ub_\nn$. The electric field in the frame of the gas is $\Eb'=\Eb+\ub_\nn\times\Bb$, and the balances of the electrons and the ions become
\begin{align}
0={}&-\ee\nee(\Eb'+\mathbf w_\ee\times\Bb)-\mee\nee\nu_{\ee\nn}\mathbf w_\ee\notag\\
&-\mee\nee\nu_{\ee\ii}(\mathbf w_\ee-\mathbf w_\ii),\label{eq:mom_e}\\
0={}&\ee\nee(\Eb'+\mathbf w_\ii\times\Bb)-\mi\nee\nu_{\ii\nn}\mathbf w_\ii\notag\\
&+\mee\nee\nu_{\ee\ii}(\mathbf w_\ee-\mathbf w_\ii).\label{eq:mom_i}
\end{align}
The two balances are linear in the drifts and are inverted for the current.
For the axial field the component along $\Bb$ decouples from the $(r,\theta)$
plane. In that plane we write the drift and the field as the complex numbers
$U_j=w_{jr}+i\,w_{j\theta}$ and $\mathcal E'=E'_r+i\,E'_\theta$, with $i$ the
imaginary unit, so that $\mathbf w_j\times\hat{\mathbf z}$ becomes $-i\,U_j$.
Divided by $m_j\nee$, \eqref{eq:mom_e} and \eqref{eq:mom_i} are then
\begin{align}
\nu_\ee(1-i\,\beta_\ee)\,U_\ee-\nu_{\ee\ii}\,U_\ii&=-\frac{\ee}{\mee}\,\mathcal E',\notag\\
\nu_\ii(1+i\,\beta_\ii)\,U_\ii-\nu_{\ii\ee}\,U_\ee&=\frac{\ee}{\mi}\,\mathcal E',
\label{eq:complex}
\end{align}
with the gyrofrequencies $\Omega_j=\ee B_0/m_j$, the momentum-loss
frequencies $\nu_\ee=\nu_{\ee\nn}+\nu_{\ee\ii}$ and
$\nu_\ii=\nu_{\ii\nn}+\nu_{\ii\ee}$ of the two species, and the Hall parameters $\beta_j=\Omega_j/\nu_j$.
This system has the
determinant
\begin{equation}
\Delta=\nu_\ee\nu_\ii\,(1-i\,\beta_\ee)(1+i\,\beta_\ii)-\nu_{\ee\ii}\nu_{\ii\ee}.
\label{eq:calD}
\end{equation}
Its solution gives the currents of the two species,
$\Jb_\ee=-\ee\nee\mathbf w_\ee$ and $\Jb_\ii=\ee\nee\mathbf w_\ii$, as
$J_{\ee r}+i\,J_{\ee\theta}=(\sigma_{\ee\mathrm P}+i\,\sigma_{\ee\mathrm H})\,\mathcal E'$
and
$J_{\ii r}+i\,J_{\ii\theta}=(\sigma_{\ii\mathrm P}-i\,\sigma_{\ii\mathrm H})\,\mathcal E'$,
with the Pedersen (subscript P) and Hall (subscript H) conductivities of the two species
\begin{align}
\sigma_{\ee\mathrm P}+i\,\sigma_{\ee\mathrm H}&=\frac{\nee\ee^2}{\mee}\,
\frac{\nu_{\ii\nn}+i\,\beta_\ii\nu_\ii}{\Delta},\label{eq:sigma_e}\\
\sigma_{\ii\mathrm P}-i\,\sigma_{\ii\mathrm H}&=\frac{\nee\ee^2}{\mi}\,
\frac{\nu_{\ee\nn}-i\,\beta_\ee\nu_\ee}{\Delta}.\label{eq:sigma_i}
\end{align}
In the total current $\Jb=\Jb_\ee+\Jb_\ii$ the magnetic terms of the two
numerators cancel, because $\beta_\ii\nu_\ii/\mee=\beta_\ee\nu_\ee/\mi$, and the response
of the plane is one complex conductivity,
\begin{gather}
J_r+i\,J_\theta=\frac{N}{\Delta}\,\mathcal E',\label{eq:sigma_tot}\\
N=\nee\ee^2\Big(\frac{\nu_{\ee\nn}}{\mi}+\frac{\nu_{\ii\nn}}{\mee}\Big).\notag
\end{gather}
Along the field the magnetic terms vanish, and the same expressions hold with
$\beta_\ee=\beta_\ii=0$.

With the real and imaginary parts of the determinant,
$\Delta=\nu_\ee\nu_\ii\,[\chi+\beta_\ee\beta_\ii-i\,(\beta_\ee-\beta_\ii)]$, where $\chi=1-\nu_{\ee\ii}\nu_{\ii\ee}/(\nu_\ee\nu_\ii)$, this is Ohm's law
$\Jb=\bm\sigma\cdot\Eb'$ with the conductivity tensor, in $(r,\theta,z)$,
\begin{equation}
\bm\sigma=\begin{pmatrix}\sP&-\sH&0\\ \sH&\sP&0\\ 0&0&\spar\end{pmatrix},
\label{eq:tensor}
\end{equation}
\begin{gather}
\spar=\frac{N}{\chi\,\nu_\ee\nu_\ii},\label{eq:sigmas}\\
\sP=\spar\,\frac{\chi\,(\chi+\beta_\ee\beta_\ii)}{(\chi+\beta_\ee\beta_\ii)^2+(\beta_\ee-\beta_\ii)^2},\notag\\
\sH=\spar\,\frac{\chi\,(\beta_\ee-\beta_\ii)}{(\chi+\beta_\ee\beta_\ii)^2+(\beta_\ee-\beta_\ii)^2}.\notag
\end{gather}
Here $\sP$, $\sH$ and $\spar$ are the Pedersen, Hall and parallel conductivities of the plasma.
The factor $\chi$ is 1 without electron-ion collisions, where these are the conductivities in common use for weakly ionized plasmas, and goes to 0 in the fully ionized limit.
The species conductivities combine to these:
$\sP=\sigma_{\ee\mathrm P}+\sigma_{\ii\mathrm P}$,
$\sH=\sigma_{\ee\mathrm H}-\sigma_{\ii\mathrm H}$ and
$\spar=\sigma_{\ee\parallel}+\sigma_{\ii\parallel}$, where
$\sigma_{j\parallel}$ is \eqref{eq:sigma_e} or \eqref{eq:sigma_i} with
$\beta_\ee=\beta_\ii=0$.

With this tensor, $\Jb=\bm\sigma\cdot\Eb'$ is the generalized Ohm's law of
a partially ionized plasma with the Hall and ion-slip
terms~\cite{MitchnerKruger1973,Song2001}. The tensor is the exact solution of
the balances \eqref{eq:mom_e} and \eqref{eq:mom_i}.
It reduces to the conductivities that Ref.~\cite{Song2001} gives in the frame of the gas when terms of order $\mee\nu_{\ee\nn}/(\mi\nu_{\ii\nn})$ are neglected.
In the fully ionized
limit, $\nu_{\ee\nn},\nu_{\ii\nn}\to0$, it gives the conductivity of a fully
ionized plasma, $\spar\to\nee\ee^2/(\mee\nu_{\ee\ii})$.
The conductivities that the two species would have separately, each with $\nu_{\ee\ii}$ or $\nu_{\ii\ee}$ in its collision frequency, add up to twice this value.

In axisymmetry, with $E_\theta=0$ and $\Bb=B_0\hat{\mathbf z}$, the
field in the frame of the gas is
\begin{equation}
\Eb'=\Eb+\ub_\nn\times\Bb=
\begin{pmatrix}E_r+u_{\nn\theta}B_0\\ -u_{\nn r}B_0\\ E_z\end{pmatrix}
\equiv\begin{pmatrix}E'_r\\ E'_\theta\\ E_z\end{pmatrix}.
\label{eq:Eprime}
\end{equation}
With this field Ohm's law $\Jb=\bm\sigma\cdot\Eb'$ is
\begin{equation}
\begin{pmatrix}J_r\\ J_\theta\\ J_z\end{pmatrix}=
\begin{pmatrix}\sP&-\sH&0\\ \sH&\sP&0\\ 0&0&\spar\end{pmatrix}
\begin{pmatrix}E_r+u_{\nn\theta}B_0\\ -u_{\nn r}B_0\\ E_z\end{pmatrix}.
\label{eq:ohm_matrix}
\end{equation}
In components, with one term of $J_r$ left out,
\begin{align}
J_r&=\sP(E_r+u_{\nn\theta}B_0), \label{eq:Jr}\\
J_\theta&=\sH(E_r+u_{\nn\theta}B_0){}-\sP u_{\nn r}B_0, \label{eq:Jt}\\
J_z&=\spar E_z. \label{eq:Jz}
\end{align}
We leave out the term $\sH u_{\nn r}B_0$ of $J_r$, the Hall current of the
radial gas flow, because it makes the explicit coupling to the gas
numerically unstable. It can be left
out because it is small. The velocity of the gas is dominated by its rotation,
and between 2.5 and 3.5~ms $\int|\sH u_{\nn r}B_0|\,\dd V$ is 6\,\% of the
rotation term of $J_r$, $\int|\sP u_{\nn\theta}B_0|\,\dd V$, on average over
the five simulations (4 to 8\,\%). The term of the radial flow in $J_\theta$ is
kept because it is the brake on the radial motion of the gas: its force on the
gas, $-\sP B_0^2u_{\nn r}$, opposes that motion
(Section~\ref{sec:neutral}), and simulations without it are unsteady at
1.25~kA and above. The term left out of $J_r$ exerts an azimuthal force and is
no brake on that motion.

The potential follows from current continuity, $\nabla\cdot\Jb=0$. With
$\Eb=-\nabla\phi$ and the current density components \eqref{eq:Jr} to \eqref{eq:Jz}, the equation
that is solved is
\begin{equation}
\nabla\cdot(\bm\sigma\cdot\nabla\phi)=\nabla\cdot(\sP\,\ub_\nn\times\Bb).
\label{eq:phi}
\end{equation}

\paragraph{Boundary conditions.}
The discharge current is imposed on the cathode tip as a Gaussian
current-density profile,
$-\spar\partial_z\phi=J_z^\mathrm{c}(r)\propto I(t)\exp(-r^2/2\sigma_r^2)$, with $J_z^\mathrm{c}$ the axial current density on the tip and the width
$\sigma_r=r_\mathrm{c}/3$, where $r_\mathrm{c}=2.75$~cm is the cathode radius,
the profile being normalized so that it carries the current $I(t)$. The prescribed profile
simplifies the setup: the cold cathode of the experiment probably carried its
current in cathode spots, whose physics is outside the scope of this work.
The vessel wall, at the outer radius $r=r_{\max}=18$~cm, is held at $\phi=0$. The cathode side,
the end wall and the axis carry no normal current, and the midplane is a
symmetry plane.

\paragraph{Current partition and heating.}
Ohm's law gives the current of each species, $\Jb=\Jb_\ee+\Jb_\ii$, with
the conductivities \eqref{eq:sigma_e} and \eqref{eq:sigma_i}. Since
$\Eb=\Eb'-\ub_\nn\times\Bb$, the electrical power is
\begin{equation}
\Jb\cdot\Eb=\Jb\cdot\Eb'+\ub_\nn\cdot(\Jb\times\Bb).
\label{eq:power}
\end{equation}
The first term is the Joule heating, the frictional dissipation of
\eqref{eq:mom_e} and \eqref{eq:mom_i}. It is divided between the species with
their currents,
\begin{equation}
q_{\Omega,j}=\Jb_j\cdot\Eb'.
\label{eq:joule}
\end{equation}
The electrical work is thus put either into Joule heating or, through the
second term, into the kinetic energy of the gas and its viscous dissipation.

\subsection{Collision frequencies}
\label{sec:collisions}

The electron-neutral collision frequency is
$\nu_{\ee\nn}=n_\nn k_\mathrm{el}(\Te)$, where $k_\mathrm{el}$ is the Maxwellian
rate coefficient of e-H elastic momentum transfer, computed from the cross section of the Morgan
database~\cite{MorganLXCat,Pitchford2017,Pancheshnyi2012,Carbone2021}, used for both isotopes. The
electron-ion collision frequency is the Braginskii
value~\cite{Braginskii1965},
\begin{gather}
\nu_{\ee\ii}=\frac{4\sqrt{2\pi}\,\nee\ee^4\ln\Lambda}{3(4\pi\varepsilon_0)^2\mee^{1/2}(\kB\Te)^{3/2}},\\
\ln\Lambda=\ln\!\big[\max\big(12\pi\nee\lambda_\mathrm{D}^3,\,2\big)\big],
\qquad\lambda_\mathrm{D}=\Big(\frac{\varepsilon_0\kB\Te}{\nee\ee^2}\Big)^{1/2},
\end{gather}
with $\varepsilon_0$ the vacuum permittivity, and the Coulomb logarithm $\ln\Lambda$ and the Debye length $\lambda_\mathrm{D}$ evaluated locally.

For the ions, resonant charge exchange with the neutrals dominates. In such a
collision the ion and the atom exchange their identities, so the new ion has
the velocity of the atom. For equal masses the relative velocity
$\mathbf g$ of an ion and an atom of species $s$ is Maxwellian at the
temperature $T_\mathrm{eff}=\Ti+T_s$, and the friction and the heat exchange
are both set by one transport coefficient,
\begin{equation}
k_\mathrm{eff}(T_\mathrm{eff})=\frac{m_\mathrm{p}}{3\kB T_\mathrm{eff}}\,
\big\langle\sigma_\mathrm{cx}(g)\,g^3\big\rangle,
\label{eq:keff}
\end{equation}
where $m_\mathrm{p}$ is the proton mass, $g=|\mathbf g|$, the average is over that Maxwellian and $\sigma_\mathrm{cx}$ is the
cross section of
$\mathrm{H}^++\mathrm{H}(1s)\rightarrow\mathrm{H}(1s)+\mathrm{H}^+$, reaction
H.1~3.1.8 of AMJUEL~\cite{ReiterAMJUEL}: the total charge-exchange cross
section of Janev \textit{et al.}~\cite{Janev1987}. The charge-exchange
frequency of the ions with species $s$ is
\begin{equation}
\nu_{\mathrm{cx},s}=n_s\,k_\mathrm{eff}(\Ti+T_s)\,\Big(\frac{m_\mathrm{p}}{\mi}\Big)^{1/2},
\label{eq:nucx}
\end{equation}
with $n_s$ the number density of that species. The coefficient is evaluated for the H$^+$ + H pair and scaled with
$(m_\mathrm{p}/\mi)^{1/2}$ to the mean ion mass, that is, the cross section is
taken at the same collision energy for all isotopes. The drift of the ions
relative to the gas is not included in the relative speed; in the simulations
it is below a fifth of the thermal speed $[\kB(\Ti+T_s)/\mi]^{1/2}$ where 90\,\%
of the radial current flows. Species $s$ exerts
on the ions the friction force
$\nee m_s\nu_{\mathrm{cx},s}(\ub_s-\ub_\ii)$, with $m_s$ the atomic mass of species $s$ and $\ub_\ii$ the velocity of the ions, so the momentum-transfer
frequency of the single ion fluid of mass $\mi$ is
\begin{equation}
\nu_{\ii\nn}=\sum_s\nu_{\ii s},\qquad\nu_{\ii s}=\frac{m_s}{\mi}\,\nu_{\mathrm{cx},s}.
\label{eq:nuin}
\end{equation}

\subsection{Plasma density and electron energy}
\label{sec:plasma}

The quasineutral density evolves as
\begin{equation}
\partial_t\nee=\nabla\cdot(\bm D_a\cdot\nabla\nee)+k_\mathrm{ion}n_\nn\nee-k_\mathrm{rec}\nee^2,
\label{eq:ne}
\end{equation}
where $n_\nn$ is the total neutral density, $k_\mathrm{ion}(\Te,\nee)$ and
$k_\mathrm{rec}(\Te,\nee)$ the effective collisional-radiative ionization and
recombination coefficients, the latter for radiative and three-body
recombination together (Section~\ref{sec:atomic}),
and $\bm D_a$ the ambipolar diffusion tensor, with the components $D_{a\parallel}$ along $\Bb$ and
$D_{a\perp}$ across it,
\begin{equation}
D_{ak}=\frac{\mu_{\ii k}\mu_{\ee k}}{\mu_{\ii k}+\mu_{\ee k}}\frac{\kB(\Te+\Ti)}{\ee},\quad k={\parallel},{\perp}.
\end{equation}
The species mobilities are $\mu_{\ee\parallel}=\ee/(\mee\nu_\ee)$,
$\mu_{\ii\parallel}=\ee/[\mi(\nu_{\ii\nn}+\nu_{\ii\ee})]$ and
$\mu_{j\perp}=\mu_{j\parallel}/(1+\beta_j^2)$.

As stated in Section~\ref{sec:assumptions}, \eqref{eq:ne} contains no advection by the
ion drift. The plasma is therefore not carried by the cross-field ion current;
its distribution results from local ionization and recombination balanced
against ambipolar diffusion. We return to this point in Section~\ref{sec:validity-plasma}.

\paragraph{Wall losses.}
On the vessel wall and on the insulating end wall, the ion flux at the sheath
edge, $\nee(\kB\Te/\mi)^{1/2}$, is applied in series with the diffusive flux over half
a grid cell, so the wall flux is limited by whichever is smaller. There is no ion
sink at the cathode. Ions lost at walls return to the neutrals as atoms at the
local gas temperature and velocity. The ionization sink, the recombination
source and the wall recycling are divided between H and D by the local neutral
mole fraction, as the ion mass \eqref{eq:mi} assumes; the single ion fluid does not
carry the isotope identity of the ions it transports. The ratio of D to H in the whole inventory, gas
and ions together, is therefore not conserved exactly: it differs from that of the fill by at most
$3\times10^{-3}$ between 2.5 and 3.5~ms in the five simulations; isotope-resolved ion transport, and the radial ion wind that could
separate the ion species, are not represented.

The electron temperature follows from the electron energy balance,
\begin{equation}
\begin{split}
\tfrac32\nee\kB\partial_t\Te={}&\nabla\cdot(\bm\kappa_\ee\cdot\nabla\Te)+q_{\Omega,\ee}\\
&-n_\nn\nee\big[K_\mathrm{ion}+\tfrac32\kB\Te\,k_\mathrm{ion}\big]\\
&-\nee^2\big[K_\mathrm{rec}-E_\mathrm{iz}k_\mathrm{rec}\big]\\
&-3\nee\kB\sum_s\frac{\mee}{m_s}\nu_{\ee s}(\Te-T_s)\\
&-3\frac{\mee}{\mi}\nee\kB\nu_{\ee\ii}(\Te-\Ti),
\end{split}
\label{eq:Te}
\end{equation}
where $\bm\kappa_\ee$ is the anisotropic electron thermal conductivity, with the
components $\kappa_{\ee\parallel}=3.16\,\nee\kB^2\Te/(\mee\nu_\ee)$ along $\Bb$, the
Braginskii form~\cite{Braginskii1965} with $\nu_\ee$ in place of the electron-ion
frequency, and $\kappa_{\ee\perp}=\kappa_{\ee\parallel}/(1+\beta_\ee^2)$ across it, with
$\nu_\ee=\nu_{\ee\nn}+\nu_{\ee\ii}$, and $q_{\Omega,\ee}$ is the electron Joule heating
\eqref{eq:joule}. The second line is the energy exchanged in ionization.
$K_\mathrm{ion}(\Te,\nee)$ is the energy-weighted counterpart of $k_\mathrm{ion}$, in
J\,m$^3$\,s$^{-1}$: it gives the energy exchanged where $k_\mathrm{ion}$ gives the number of events.
$K_\mathrm{ion}\,n_\nn\nee$ is the power per unit volume that the
electrons spend on ionization: the ionization energy $E_\mathrm{iz}=13.6$~eV of each atom ionized
and the radiation that escapes after excitation. The term in
$\tfrac32\kB\Te$ is the thermal energy that the other electrons give to each
new electron to bring it to $\Te$. The third line is the energy exchanged in recombination, with
$K_\mathrm{rec}(\Te,\nee)$ the energy-weighted counterpart of $k_\mathrm{rec}$.
$K_\mathrm{rec}\,\nee^2$ is the power radiated per unit volume, and
$E_\mathrm{iz}k_\mathrm{rec}\nee^2$ is the ionization energy that
recombination returns to the electrons per unit volume and time. A recombining electron takes its
thermal energy $\tfrac32\kB\Te$ with it, which leaves $\Te$ unchanged.
Both coefficients come from the
collisional-radiative model that gives $k_\mathrm{ion}$ and $k_\mathrm{rec}$
(Section~\ref{sec:atomic}), so the excited states and the radiation are not
tracked as separate unknowns. The last two lines are the elastic energy exchange
with each neutral species, with $\nu_{\ee s}=(n_s/n_\nn)\nu_{\ee\nn}$, and with the ions.

\paragraph{Sheath heat losses.}
The sheath heat flux at the vessel wall and the end wall is
$\delta_T\nee(\kB\Te/\mi)^{1/2}\kB\Te$ with the sheath heat transmission factor $\delta_T=7$, applied in series with
conduction over half a cell. Since no ion sink is applied at the cathode, no
sheath heat sink is applied there either. The midplane is adiabatic by symmetry.

\subsection{Atomic data}
\label{sec:atomic}

The plasma species are electrons, atomic ions and ground-state atoms; the same
atomic data are used for H and D, which differ only by their masses.

\paragraph{Ionization and recombination.}
The coefficients $k_\mathrm{ion}$ and $k_\mathrm{rec}$ are the effective
collisional-radiative rate coefficients of the AMJUEL
database~\cite{ReiterAMJUEL,SawadaFujimoto1995}: reaction H.4~2.1.5 for
$\mathrm{e}+\mathrm{H}(1s)\rightarrow2\mathrm{e}+\mathrm{H}^+$ and reaction
H.4~2.1.8 for $\mathrm{H}^++\mathrm{e}\rightarrow\mathrm{H}(1s)$, which
contains radiative and three-body recombination together.
The fits are valid for $0.1\le\Te\le2\times10^4$~eV
and $10^{14}\le\nee\le10^{22}$~m$^{-3}$; outside this range the value at the
boundary is used.

\paragraph{Electron energy.}
The energy-weighted coefficients $K_\mathrm{ion}$ and $K_\mathrm{rec}$ of
\eqref{eq:Te} are reactions H.10~2.1.5 and H.10~2.1.8 of the same
database, with the same range.

All four coefficients assume that the line and continuum radiation escapes:
they are those of an optically thin plasma, and the trapping of the Lyman
lines is not included.

\subsection{Ion temperature and azimuthal momentum}
\label{sec:ions}

The ion temperature follows from the ion energy balance,
\begin{equation}
\begin{split}
\tfrac32n_\ii\kB\partial_t\Ti={}&\nabla\cdot(\bm\kappa_\ii\cdot\nabla\Ti)+\Big(1-\sum_sf_s\Big)q_{\Omega,\ii}\\
&+3\frac{\mee}{\mi}n_\ii\nu_{\ee\ii}\kB(\Te-\Ti)\\
&+\tfrac32n_\ii\kB\sum_s\nu_{\mathrm{cx},s}(T_s-\Ti)\\
&+\tfrac32\kB\,k_\mathrm{ion}\nee\sum_sn_s(T_s-\Ti),
\end{split}
\label{eq:Ti}
\end{equation}
where $\bm\kappa_\ii$ is the ion thermal conductivity, with the components $\kappa_{\ii\parallel}=n_\ii\kB^2\Ti/[\mi(\nu_\ii+\nu_{\ii\ii})]$ along $\Bb$ and
$\kappa_{\ii\perp}=\kappa_{\ii\parallel}/(1+\beta_\ii^2)$ across it, and $\nu_{\ii\ii}$ is the
Spitzer ion-ion frequency with the Coulomb logarithm of $\nu_{\ee\ii}$
(Section~\ref{sec:collisions}). The walls are adiabatic for
the ions. The ion Joule
heating $q_{\Omega,\ii}$ includes the frictional dissipation of the ion drift
relative to the gas; the fractions $f_s$ of it given directly to the neutral
species are defined in Section~\ref{sec:neutral}. The third line is the heat
exchanged with the neutral species by charge exchange: the ion and the atom
exchange their velocities, so the temperatures relax at the charge-exchange
frequency $\nu_{\mathrm{cx},s}$ itself \eqref{eq:nucx}, without the mass factor
of elastic collisions. The last line is the enthalpy carried by the reactions:
an ion created by ionization appears at the temperature of the atom it came
from. The atoms created by recombination appear at $\Ti$
(Section~\ref{sec:neutral}).

The ion rotation is resolved as a separate unknown:
\begin{equation}
\mi n_\ii\partial_tu_{\ii\theta}=\hat{\bm\theta}\cdot\big[\Jb_\ii\times\Bb-\mi n_\ii\nu_\ii(\ub_\ii-\bar{\ub})+\eta_0\nabla^2\ub_\ii\big],
\label{eq:uitheta}
\end{equation}
where $\hat{\bm\theta}$ is the azimuthal unit vector, $\nu_\ii=\nu_{\ii\nn}+\nu_{\ii\ee}$,
$\bar{\ub}=(\nu_{\ii\nn}\ub_\nn+\nu_{\ii\ee}\ub_\ee)/\nu_\ii$ with the azimuthal velocity of the electrons $u_{\ee\theta}$ given below, and the ion viscosity
$\eta_0=0.96\,n_\ii\kB\Ti\,\nu_\ii^{-1}/[1+(2\beta_\ii)^2]$ is the magnetized
Braginskii form~\cite{Braginskii1965} with the ion-neutral and ion-electron
frequencies in place of the ion-ion frequency, an approximate interpolation
for an ion fluid dominated by collisions with neutrals. The ions and the neutral gas are
separate species, each with its own viscosity; the ion viscosity is included
but is below 1\,\% of the gas viscosity in most of the plasma and at most 14\,\%
of it. The ions do not slip at the walls
and at the cathode, and $\partial_zu_{\ii\theta}=0$ on the midplane. In the
drag-dominated limit, and for $E'_\theta=0$, the equation reduces to the
azimuthal component of \eqref{eq:mom_i}; where the ion-electron friction is
also negligible it gives the ion Hall drift relative to the gas,
\begin{equation}
u_{\ii\theta}-u_{\nn\theta}=-\frac{\beta_\ii^2}{1+\beta_\ii^2}\frac{E'_r}{B_0},
\end{equation}
consistent with $J_{\ii r}=\sigma_{\ii\mathrm P}E'_r$. The electron azimuthal
velocity follows from the algebraic electron balance,
\begin{equation}
u_{\ee\theta}=\frac{\nu_{\ee\ii}u_{\ii\theta}+\nu_{\ee\nn}u_{\nn\theta}-J_{\ee r}B_0/(\mee\nee)}
{\nu_{\ee\ii}+\nu_{\ee\nn}}.
\end{equation}

\subsection{Neutral species}
\label{sec:neutral}

Each neutral species has the mass density $\rho_s=m_sn_s$ and the pressure $p_s=n_s\kB T_s$ and obeys
\begin{gather}
\partial_t\rho_s+\nabla\cdot(\rho_s\ub_s)=S_s,\label{eq:neutral_mass}\\
\begin{split}
\partial_t(\rho_s\ub_s)&+\nabla\cdot(\rho_s\ub_s\ub_s)\\
&=-\nabla p_s+\nabla\cdot\bm\tau_s+\mathbf F^\mathrm{pl}_s+\mathbf C_s+S_s\ub_s,
\end{split}\\
\begin{split}
\partial_t(\rho_sc_vT_s)&+\nabla\cdot(\rho_sc_vT_s\ub_s)\\
&=-p_s\nabla\cdot\ub_s+\bm\tau_s\!:\!\nabla\ub_s+\nabla\cdot(\kappa_s\nabla T_s)\\
&\quad+Q^\mathrm{pl}_s+Q^\mathrm{C}_s+S_sc_vT_s,
\end{split}\label{eq:neutral_E}
\end{gather}
where $S_s$ is the net source of species $s$ from recombination, wall
recycling and ionization; $\bm\tau_s$ is the Newtonian stress with partial
viscosity $\mu_s$; $\mathbf F^\mathrm{pl}_s$ is the force exerted by the plasma
and $\mathbf C_s$ the friction and thermal force exerted by the other neutral
species (both given below); $c_v=\tfrac32\kB/m_s$ is the specific heat at constant volume; $\kappa_s$ is the partial
thermal conductivity; and $Q^\mathrm{pl}_s$ and $Q^\mathrm{C}_s$ are the heating by the
plasma and by the other neutral species. The terms $S_s\ub_s$ and $S_sc_vT_s$ are the
momentum and the thermal energy that the source carries at the velocity and the temperature of the
species.

\begin{figure}
\includegraphics[width=\columnwidth]{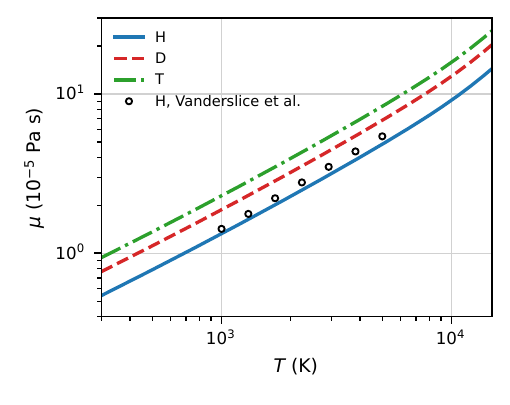}
\caption{\label{fig:visc}Viscosity of atomic H, D and T in the first Chapman-Enskog
approximation, with the H-H collision integrals of Ref.~\cite{Bruno2010}
($\mu_s\propto m_s^{1/2}$). Circles: atomic hydrogen, Vanderslice
\emph{et al.}~\cite{Vanderslice1962} as fitted in Ref.~\cite{Svehla1995}.}
\end{figure}

\paragraph{Transport coefficients.}
The pure-species viscosity and conductivity are the first Chapman-Enskog
approximation~\cite{ChapmanCowling1970,Hirschfelder1964},
\begin{equation}
\mu_s=\frac{5}{16}\frac{\sqrt{\pi m_s\kB T}}{\pi\,\overline{\Omega}^{(2,2)}(T)},\qquad
\kappa_s=\frac{15}{4}\frac{\kB}{m_s}\mu_s,
\label{eq:transport}
\end{equation}
where $T$ is the temperature of the gas mixture and $\pi\overline{\Omega}^{(l,s)}$ is the thermally averaged transport cross
section of order $(l,s)$. For the hydrogen isotopes a Lennard-Jones model is a poor
representation of two ground-state atoms, which interact on the singlet
$X\,^1\Sigma_g^+$ and triplet $b\,^3\Sigma_u^+$ curves of H$_2$ with statistical
weights $1/4$ and $3/4$. We therefore use the collision integrals
$\overline{\Omega}^{(1,1)}$, $\overline{\Omega}^{(1,2)}$,
$\overline{\Omega}^{(1,3)}$ and $\overline{\Omega}^{(2,2)}$ of H-H fitted by
Bruno \emph{et al.}~\cite{Bruno2010} to accurate calculations on the ab initio
potentials, valid from 50 to 50\,000~K. In the Born-Oppenheimer picture the same
integrals apply to H, D and T, so that $\mu_s\propto m_s^{1/2}$.
The viscosities of H, D and T are shown in Fig.~\ref{fig:visc}. The
integrals are tabulated once and interpolated in $\ln T$ during the run. The
coefficients are distributed between species with Wilke's
rule~\cite{Wilke1950,MasonSaxena1958}.
An empirical Knudsen factor of Bosanquet type~\cite{Wang2016}, $(1+\Kn_s)^{-1}$, is then applied, with $\Kn_s=\lambda_s/r_{\max}$
and a mean free path $\lambda_s$ that includes self and cross collisions,
evaluated with the same collision integrals; it
is not a kinetic closure. In the simulations at the current peak it lowers the
coefficients by 1 to 2\,\% at the inner sampling radius and, on the axis at the
midplane, by 15 to 25\,\% for H and 15 to 31\,\% for D
(Section~\ref{sec:validity-neutral}).

\paragraph{Inter-species coupling.}
The species exchange momentum through binary friction, which conserves total
momentum. For the H/D pair,
\begin{gather}
\mathbf C_\mathrm{H}=-\mathbf C_\mathrm{D}=-\rho_\mathrm{H}\nu_\mathrm{HD}(\ub_\mathrm{H}-\ub_\mathrm{D}),\\
\nu_\mathrm{HD}=\frac43\,\frac{m_\mathrm{D}}{m_\mathrm{H}+m_\mathrm{D}}\,n_\mathrm{D}\,\pi\overline{\Omega}^{(1,1)}(T)\Big(\frac{8\kB T}{\pi\mu_\mathrm{HD}}\Big)^{1/2}.
\end{gather}
Here $\mu_\mathrm{HD}=m_\mathrm{H}m_\mathrm{D}/(m_\mathrm{H}+m_\mathrm{D})$ is the reduced mass of the pair.
With this Chapman-Enskog momentum-transfer frequency the pair coefficient is
symmetric, $\rho_\mathrm{H}\nu_\mathrm{HD}=\rho_\D\nu_\mathrm{DH}=K_\mathrm{HD}=n_\mathrm{H}n_\D\kB T/(nD_\mathrm{HD})$,
with $n=n_\mathrm{H}+n_\D$ the total number density, $\nu_\mathrm{DH}$ the frequency of D on H and $D_\mathrm{HD}$ the first Chapman-Enskog binary diffusion coefficient, so that the
steady slip reproduces the binary diffusion flux. The kinetic energy dissipated
by the friction heats the two species in the inverse ratio of their masses. The species also exchange heat,
\begin{equation}
Q^\mathrm{C}_{\mathrm{H}}\supset3\frac{m_\mathrm{H}}{m_\mathrm{H}+m_\mathrm{D}}\,n_\mathrm{H}\nu_\mathrm{HD}\kB(T_\mathrm{D}-T_\mathrm{H}),
\end{equation}
where $\supset$ marks a term contained in the quantity on its left, with the opposite term in $Q^\mathrm{C}_{\mathrm{D}}$, and $\nu_\mathrm{HD}$
evaluated at the reduced temperature
$(m_\mathrm{D}T_\mathrm{H}+m_\mathrm{H}T_\mathrm{D})/(m_\mathrm{H}+m_\mathrm{D})$.

Thermal diffusion enters as a thermal force in the same exchange,
\begin{equation}
\mathbf C_\mathrm{D}\supset-\alpha_T\,x_\mathrm{H}x_\mathrm{D}\,n\kB\nabla T,
\end{equation}
with the opposite force on H, $\alpha_T$ the thermal-diffusion factor, $x_s$ the mole fractions, $n=n_\mathrm{H}+n_\mathrm{D}$
and $T$ the mixture temperature. With the friction above, the steady slip it
drives reproduces the Chapman-Enskog thermal-diffusion flux, and its no-flux
equilibrium, $\mathrm{d}\ln(x_\mathrm{D}/x_\mathrm{H})=-\alpha_T\,\mathrm{d}\ln T$, does
not depend on the friction coefficient. The thermal-diffusion factor is the
first Chapman-Enskog approximation for a quasi-isotopic
pair~\cite{ChapmanCowling1970,Hirschfelder1964},
$\alpha_T=\tfrac{105}{118}R_T\,(m_\mathrm{D}-m_\mathrm{H})/(m_\mathrm{D}+m_\mathrm{H})$,
with the factor $R_T=\tfrac{118}{105}\,15(2A^*+5)(6C^*-5)/[2A^*(16A^*-12B^*+55)]$ and $A^*$,
$B^*$, $C^*$ the ratios of the collision integrals above. For
H/D, $\alpha_T=0.17$ at 300~K and 0.08 at 3000~K. It changes sign near 6000~K
and is $-0.09$ at $10^4$~K,
because the deep singlet well makes the
interaction softer than that of Maxwell molecules at these energies. The
sign change follows from the tabulated integrals of Ref.~\cite{Bruno2010};
it rests, however, on the first approximation for a
quasi-isotopic pair, which is less accurate for the 2:1 mass ratio of H/D.

There is no explicit diffusion flux in the model. Separation arises from the
slip between the species, given by their momentum equations. In the friction-dominated limit
this slip reduces to the Maxwell-Stefan form of ordinary, pressure and
thermal diffusion, whose zero is the local equilibrium \eqref{eq:alpha}
corrected by the thermal term.

\paragraph{Plasma coupling.}
\label{sec:plcoupling}
In the azimuthal direction each species is dragged toward the ion and electron
velocities,
\begin{equation}
\begin{split}
F^\mathrm{pl}_{s\theta}={}&\mi n_\ii\nu_{\ii s}(u_{\ii\theta}-u_{s\theta})\\
&+\mee\nee\nu_{\ee s}(u_{\ee\theta}-u_{s\theta}),
\end{split}
\end{equation}
where $\nu_{\ii s}=(m_s/\mi)\nu_{\mathrm{cx},s}$ is the momentum-transfer frequency
of the ions with species $s$ \eqref{eq:nuin}.

In the radial direction the Hall current exerts the force $J_\theta B_0$
\eqref{eq:Jt} on the plasma. This force is inward, since $E'_r<0$. It is passed
to the gas as a force density shared between the species in proportion to
their mass density. No other $(r,z)$ plasma force acts on the gas.

The collisional heating of each species by the plasma is
\begin{equation}
\begin{split}
Q^\mathrm{pl}_s={}&3\kB\nee\frac{\mee}{m_s}\nu_{\ee s}(\Te-T_s)
+\tfrac32\kB\nee\nu_{\mathrm{cx},s}(\Ti-T_s)\\
&+\tfrac32\kB\,S^\mathrm{rec}_s(\Ti-T_s){}+f_sq_{\Omega,\ii},
\end{split}
\end{equation}
where $S^\mathrm{rec}_s$ is the number of atoms of species $s$ created by
volume recombination per unit volume and time, which appear at $\Ti$. The
elastic and charge-exchange terms, summed over species, are removed from the
electron and ion energy
equations, so the exchange terms cancel pairwise; since the electron, ion and
neutral blocks are advanced at different stages of the time step, the exchange
conserves energy only to the splitting error. The ion Joule heating
$q_{\Omega,\ii}=\Jb_\ii\cdot\Eb'$ \eqref{eq:joule}
is the dissipation of the
ion-neutral friction; species $s$ receives the share
$f_s=\tfrac12\,\nu_{\ii s}/\sum_q\nu_{\ii q}$ of it, half of the dissipation of
its own friction, as for a resonant pair of equal masses, and the ions
keep the rest.

\paragraph{Boundary conditions and chemistry.}
The vessel is sealed. The walls, including the side of the cathode, are held
at 300~K; the cathode tip is a hot surface at 1200~K, which heats gas colder than
itself and is adiabatic for hotter gas. The walls have a velocity slip and the Smoluchowski temperature
jump~\cite{Kennard1938}; the slip on the curved walls is $-L_\mathrm{s}\partial_ru_\theta$, with $L_\mathrm{s}$ the slip length; this form omits the curvature term $u_\theta/r$ of Maxwell's condition
(Section~S1\,B of the supplementary material). The tangential momentum accommodation is unity; the
thermal accommodation coefficients of H and D on steel, 0.12 and 0.22, follow
the correlation of Song and Yovanovich~\cite{SongYovanovich1987}. The axis and the midplane are
symmetry boundaries. The ionization sink, the recombination source and the
wall recycling are divided between species by the local neutral mole
fraction.

\section{Numerical method}
\label{sec:numerics}

The model couples processes whose time scales span many orders of magnitude.
Electron-neutral collisions take about $10^{-10}$~s. Ion-neutral momentum
and energy exchange take $10^{-8}$ to $10^{-7}$~s, as do the conduction and
viscous times on the cell scale. The discharge pulse lasts several
milliseconds. In addition, the potential equation is strongly anisotropic, with
$\spar/\sP\simeq(\beta_\ee/\beta_\ii)(1+\beta_\ii^2)\sim10$ to $10^{2}$ in the
discharge and much larger outside it. Most earlier plasma-centrifuge models avoided these
difficulties by seeking steady states of reduced
equations~\cite{Wijnakker1979,VanDenBerg1982,Simpson1981b,Yue1996}. Here the
full coupled system is integrated in time over the pulse. The integration
rests on three choices. Every stiff local exchange is treated implicitly or
integrated analytically, so the macro time step is set by the evolution of the
discharge and not by collision times. The potential is obtained at every step
by a direct sparse solve, which is insensitive to the anisotropy. Only the
neutral advection is explicit, and it is subcycled at its own acoustic limit.
With a macro step of $\Delta t=1~\mu$s, the 3.5~ms simulated here take
3500 steps.

\paragraph{Discretization.}
The grid is uniform and vertex-centered, with $310\times621$ nodes and nodes
placed on the axis, the walls and the midplane (spacings $\Delta r=0.58$~mm,
$\Delta z=0.69$~mm). The axial spacing resolves the boundary layer on the end
plate with about 2.6 cells per layer thickness, the resolution at which the gas
solver was tested against OpenFOAM (Section~S1 of the supplementary material). The
equations for the potential and the plasma density, Eqs.~\eqref{eq:phi} and
\eqref{eq:ne}, and the neutral mass and momentum equations are discretized in
conservative finite-volume form. The neutral energy equation is advanced in
advective form, as an equation for $T_s$. The electron and ion energy equations and the
ion azimuthal momentum equation use centered finite differences. The code is written in Python,
with the grid kernels compiled with Numba; the array operations of the gas
update are compiled as well. The sparse systems of the potential and of the electron
temperature are factorized with the direct solver PARDISO of the Intel oneAPI
Math Kernel Library. The implicit heat conduction and viscosity of the gas are
solved by red-black successive over-relaxation, with the relaxation factor of
each solve set from the convergence rate observed in the previous one.
\paragraph{Time step.}
Each macro step is split into three blocks, advanced in sequence.

In the ion block, the collision frequencies and the conductivities are updated
from the current state. The ion temperature is then advanced by a local
implicit update of the energy exchange, followed by implicit anisotropic
conduction. Last, $u_{\ii\theta}$ is advanced implicitly, with the drag, the
magnetized viscosity and the Lorentz force $-J_{\ii r}B_0$ from the previous
potential solve.

The neutral block is advanced in four equal steps per macro step,
$\Delta t_\mathrm{g}=\Delta t/4$, with the
plasma state held fixed, because the drag on the gas beside the cathode is
stiff at the macro step.
In each of them the species are advected
together by explicit SSP-RK2 substeps~\cite{ShuOsher1988} at a CFL number of 0.5
based on $|u_n|+(\gamma\kB T_s/m_s)^{1/2}$, with $u_n$ the velocity component
along each grid direction and the ratio of specific heats $\gamma=5/3$. The friction and heat exchange
between the species, including the thermal force, are applied implicitly after
every Runge-Kutta stage, so that no stage uses the free-flight slip; the
friction time can be much shorter than a substep. The implicit $2\times2$ solve
conserves the momentum of the mixture exactly, and the kinetic energy it
removes is returned as heat.

The advective fluxes are designed for the low Mach number of most of the
meridional flow. All of them use a second-order (MUSCL) reconstruction of the
face values with the minmod limiter~\cite{LeVeque2002}. A Rusanov
flux~\cite{Rusanov1961,LeVeque2002}, whose dissipation scales with the sound
speed, diffuses composition and angular momentum at rates up to $10^{2}$ times
the physical ones. We therefore upwind mass and momentum with the flow speed
only, and add two dissipative terms in the spirit of Roe-type low-Mach
schemes~\cite{Roe1981,GuillardViozat1999}: a mass flux on jumps of the total
pressure, and a damping $-\tfrac12 f\rho_sc_s\,(u_n^{+}-u_n^{-})$ of jumps of the
normal velocity on the normal momentum, with $c_s$ the sound speed and the superscripts $+$ and $-$ for the two sides of a cell face. The factor $f=\min(1,M_f)$, where
$M_f=(u_r^2+u_z^2)^{1/2}/c_s$ is the meridional Mach number at the face, with $u_r$ and $u_z$ the radial and axial velocities, makes
the damping follow the flow speed, $\rho_s|\ub|\,\Delta u$ with $|\ub|=(u_r^2+u_z^2)^{1/2}$ and $\Delta u=u_n^{+}-u_n^{-}$, in slow meridional
flow and leaves it unchanged where $M_f\ge1$~\cite{GuillardViozat1999}. Without
it the acoustic damping weakens the slow inflow of the end-plate layer, which
carries the angular momentum to the wall (Section~S1 of the supplementary material).
Both dissipative terms act on reconstructed jumps: the jump $d_i$ at face $i$ of the
first-order form is replaced by
$d_i-\tfrac12\,\mathrm{minmod}(d_{i-1},d_i)-\tfrac12\,\mathrm{minmod}(d_i,d_{i+1})$,
so a smooth profile sees a dissipation smaller by $O(\Delta x^2)$, with $\Delta x$ the grid spacing, while an
odd-even mode, whose neighboring jumps have opposite signs, is damped in
full. The acoustic mass flux of the mixture is shared between the species by
mass fraction, so a composition gradient at uniform pressure is not diffused
numerically. The pressure enters as a face flux with a well-balanced
reconstruction of the centrifugal force: a gas in rigid rotation at uniform
temperature is an exact discrete equilibrium and sees no dissipation. The
Rusanov flux is used only during the first 50~$\mu$s, when the discharge forms.

The azimuthal equation is advanced as a conservation law for the angular
momentum density $\rho_sru_{s\theta}$, with no geometric source term, so the
advection step conserves the angular momentum of the gas. This
quantity is upwinded with the flow speed $|u_n|$ alone, with no added
dissipation. With a first-order reconstruction this leaves odd-even
oscillations of the rotation undamped where the gas is nearly at rest; they
could be damped by a stabilizing flux
\begin{equation}
-\tfrac12\,\epsilon\,f\,c_s\,\rho_s r^{2}\,\big(\omega^{+}-\omega^{-}\big)
\label{eq:stab}
\end{equation}
on the jumps of the angular velocity $\omega=u_{s\theta}/r$, which vanish for
rigid rotation. The second-order reconstruction does not need it, and its
coefficient $\epsilon$ is zero in the simulations.

The boundary rows (the outer wall, the end plate and the midplane) are ghost
rows that carry no mass: they mirror the density and the normal velocity of
the interior, the faces between them and the interior are closed to advection, and the wall conditions act
through the stress, the heat flux and the pressure on those faces.

The advection is followed by the implicit plasma drag and heat exchange, and by
implicit heat conduction and viscosity. The kinematic viscosity varies by two
orders of magnitude between the cold gas at the walls and the hot core. Each velocity
component is therefore solved with the mean viscosity of the two cells on
every face, so that the part of the viscous force that comes from the
gradient of the viscosity is inside the implicit solve; as an explicit source
it is unstable where the gas is thin and its viscosity changes steeply from
cell to cell. Of the viscous force, only the terms that couple different
velocity components are explicit.
The heat released by the viscosity is the kinetic energy that the
implicit viscous solve removes, deposited in proportion to the local
dissipation $\bm\tau\!:\!\nabla\ub$. The kinetic and thermal energy lost by the
advection substeps (numerical dissipation) is returned as heat in the same way.

The Lorentz force on the gas carries its own brake. At fixed potential
the back electromotive force (back-EMF) $u_{\nn\theta}B_0$ in $J_r=\sP(E_r+u_{\nn\theta}B_0)$ reduces the
azimuthal force $-J_rB_0$ as the gas spins up,
$\partial(-J_rB_0)/\partial u_{\nn\theta}=-\sP B_0^2$, and the radial force
$J_\theta B_0$ brakes the radial flow in the same way \eqref{eq:Jt}. The
potential is solved once per macro step $\Delta t$. Where the gas is thin the
brake is faster than that, $\sP B_0^2/\rho>1/\Delta t$ with
$\rho=\sum_s\rho_s$ the density of the gas, and a force taken from the
potential solve and applied explicitly overshoots: the rotation of the gas
and the discharge voltage alternate from one step to the next. Such a force
is therefore multiplied by
\begin{equation}
\frac{1}{1+K},\qquad K=\frac{\sP B_0^2\,\Delta t}{\rho},
\label{eq:dragK}
\end{equation}
the backward-Euler response of the gas to a force with this brake.

In every gas step the azimuthal friction force $F^\mathrm{pl}_{s\theta}$ of
Section~\ref{sec:neutral} is applied in two parts. First, each species is
relaxed implicitly toward the ion and electron velocities, which are held
fixed,
\begin{gather}
u_{s\theta}\leftarrow\frac{u_{s\theta}+(c_{\ii s}u_{\ii\theta}+c_{\ee s}u_{\ee\theta})\,\Delta t_\mathrm{g}/\rho_s}{1+a_s},\label{eq:dragrelax}\\
a_s=\frac{(c_{\ii s}+c_{\ee s})\,\Delta t_\mathrm{g}}{\rho_s},\notag
\end{gather}
with $c_{\ii s}=\mi n_\ii\nu_{\ii s}$ and $c_{\ee s}=\mee\nee\nu_{\ee s}$. This
delivers the fraction $1/(1+a_s)$ of $F^\mathrm{pl}_{s\theta}$. Second, the
rest of the friction force on the whole gas is added with the factor
\eqref{eq:dragK},
\begin{equation}
u_{s\theta}\leftarrow u_{s\theta}+\frac{\big[F^0_\theta-\sum_qF^\mathrm{pl}_{q\theta}/(1+a_q)\big]\,\Delta t_\mathrm{g}/\rho}{1+K},
\label{eq:dragfactor}
\end{equation}
where $F^0_\theta=\sum_sF^\mathrm{pl}_{s\theta}$ is evaluated once per macro
step, after the ion rotation is advanced, and held over the four gas steps.
The change of velocity is the same for every species, and it enters the
implicit viscous solve as a source. The radial force $J_\theta B_0$ is applied
in the same way, multiplied by $1/(1+K)$. The part of the Joule heating \eqref{eq:joule} that
comes from the radial flow of the gas, $\sigma_{j\mathrm P}(u_{\nn r}B_0)^2$,
is deposited with the same factor, so that it equals the kinetic energy that
this force takes from the gas. Because the term $\sH u_{\nn r}B_0$ is left out of $J_r$
(Section~\ref{sec:ohm}), the work that the Hall part of $J_\theta B_0$ does
on the radial flow is not drawn from the electrical power; averaged over the window in which the separation
is evaluated, 2.5 to 3.5~ms, it is at most
6\,\% of $\int\Jb\cdot\Eb\,\dd V$.

In the core of the plasma, in front of the cathode, the ionization
fraction is high. Ohm's law (Section~\ref{sec:ohm}), which drops the inertia of
the plasma and lets the gas take up the Lorentz force, is at its limit there.
This region is the main reason for the factor: without it, the explicit
coupling is unstable there. The bulk of the gas receives nearly all of
the Lorentz torque that acts on it. From 2.5 to 3.5~ms more than 94\,\% of the neutral gas lies where the ionization fraction is below 10\,\%. Of the Lorentz torque of the discharge,
75 to 92\,\% acts on this gas, and the scheme delivers
90 to 98\,\% of that part to it.

In the electron block, $\Te$ is advanced by a local Newton iteration on the
Joule heating, the reaction terms and the collisional exchange, followed by
implicit anisotropic conduction. The rate coefficients are evaluated at the
iterated $\Te$ and at the density of the start of the step. Ionization is
stiff in $\Te$: where it would deplete the atoms within one step, that is,
where $x=k_\mathrm{ion}\nee\Delta t$ is not small, the neutral density in the
ionization loss of \eqref{eq:Te} is replaced by its mean over the step,
$n_\nn[1-\exp(-x)]/x$, so that the energy charged is bounded by the atoms that
are present; $x$ exceeds one in at most 2\,\% of the cells at 1.55~kA and in
none at 0.76~kA. The density is advanced by an analytic
integration of the local ionization and recombination, followed by an implicit
diffusion step with the wall fluxes of Section~\ref{sec:plasma}. In the
mobilities of the ambipolar diffusion coefficients an
inertial frequency $c_\mathrm{s}/(2h)$, with the ion sound speed
$c_\mathrm{s}=[\kB(\Te+\Ti)/\mi]^{1/2}$ and $h$ the smaller of the two cell
sizes, is added to the collision frequency of both species; it bounds the
diffusion coefficients where collisions are rare. The ionization sink
and the recombination and recycling sources are passed to the neutral species.
The reaction enthalpy of \eqref{eq:Ti} and \eqref{eq:neutral_E} is applied
with them as a number-weighted mixing of the temperatures, which conserves the
energy of the heavy particles, using the net of ionization and recombination
in each cell.
The conductivity tensor is then rebuilt, and \eqref{eq:phi} is solved using harmonic face averages of $\sP$ and $\spar$, with the
current boundary condition at the cathode and the rotation EMF
$u_{\nn\theta}B_0$ evaluated with the
updated gas velocity. The
species currents and the Joule heating of Section~\ref{sec:ohm} follow from the new
potential, and they feed the next step.

\paragraph{Verification.}
The rotation that the gas reaches is limited by the angular momentum carried
to the walls. In the long vessel simulated here the side wall takes most of it
(Section~\ref{sec:results-hd}); the end plates are expected to take a larger
share in a shorter vessel, which this paper does not test. In either case the layer on the end plate drives the
meridional circulation that carries angular momentum and mixes the composition.
The gas solver was therefore verified on its own, without plasma, against exact
solutions and against an established, independently developed code,
OpenFOAM~13~\cite{Weller1998}, solving the same problems under identical
conditions: the viscous operators and the wall conditions, the conservation of
mass, angular momentum and energy, the separation of a rotating mixture and the
thermal-diffusion equilibrium, the spin-down of a rotating gas against walls at
rest, and the end-plate layer at the Mach number and resolution of the
simulations and under a hot core (Section~S1 of the supplementary material).
The scheme and the grid described above were chosen on these tests.
The headline results, summarized in Table~SI of the
supplementary material, are agreement with the exact solutions to better than
0.1\,\%, with the OpenFOAM spin-down to within 2.5\,\%, and with the OpenFOAM
end-plate layer to within 4\,\% in inflow and rotation and 2 to 6\,\% in pumping.
In an isolated update, the energy exchanges with the plasma conserve energy to
$10^{-6}$. The rate coefficients of Section~\ref{sec:atomic} and the charge-exchange
coefficient \eqref{eq:keff} reproduce reference
values of the fits to $2\times10^{-4}$ or better; in an isolated update the
density step conserves the number of heavy particles, and the reaction enthalpy
their energy, to round-off.

\section{Scaling estimates}
\label{sec:scaling}

This section collects two estimates against which the simulations are read:
the separation that a given rotation and temperature allow, and what limits the
rotation. They are estimates, not results of
the model.

\subsection{Equilibrium separation}
\label{sec:scaling-alpha}

Neglecting the secondary flow, each neutral species obeys the centrifugal
barodiffusion balance
\begin{equation}
\frac{\dd(n_s\kB T)}{\dd r}=\frac{n_sm_su_\theta^2}{r}.
\end{equation}
For two species with mass difference $\Delta m$, this gives the separation
relative to the axis,
\begin{equation}
\alpha(r)\equiv\frac{(n_2/n_1)_r}{(n_2/n_1)_0}
=\exp\!\left[\frac{\Delta m}{\kB}\int_0^r\frac{u_\theta^2}{T\,r'}\,\dd r'\right],
\label{eq:alpha}
\end{equation}
where $n_1$ and $n_2$ are the densities of the two species, species 2 is the heavier, and the subscripts $r$ and 0 denote the radius $r$ and the axis. For solid-body rotation with the angular velocity $\Omega$ at uniform
temperature, $\alpha=\exp(\Delta m\Omega^2r^2/2\kB T)$. Neither form describes the
device. The gas is held nearly at rest at the end plates and at the vessel
wall, so the
rotation is far from rigid and varies along $z$, and the end plates drive a
secondary flow that carries the species through the $(r,z)$ plane and
redistributes them, which the radial balance above ignores.
Equation~\eqref{eq:alpha} is therefore only a reference; the model
resolves the secondary flow and does not assume this equilibrium. Within the
radial balance, $\alpha$ depends on $u_\theta^2/T$, not on $u_\theta$ alone;
in the device it also depends on the secondary flow.

Between the sampling radii $R_1$ and $R_2$ the exponent is written as
\begin{equation}
\ln\alpha=S\,\frac{\Delta m\,u_{\max}^2}{\kB T},\qquad
S=\frac{T}{u_{\max}^2}\int_{R_1}^{R_2}\frac{u_\theta^2(r)}{T(r)\,r}\,\dd r,
\label{eq:S}
\end{equation}
with $u_{\max}$ the peak azimuthal speed, $T$ a representative gas temperature and $S$ a
dimensionless profile factor; between the axis and the wall at uniform temperature,
$S=1/2$ for rigid rotation and $S=4/3$ for a parabolic profile. For atomic H/D, the mean atomic mass is $m=1.5$~u and $\Delta m/m=2/3$.
The simulated separation factor between 8 and 17~cm is 60 to 64\,\% of the equilibrium value that
Eq.~\eqref{eq:alpha} gives with the simulated rotation and gas temperature on the midplane. It is lower than the
equilibrium estimate because the simulation includes the secondary flow driven by the end plates, which mixes the outer
layer (Section~\ref{sec:scaling-mixing}).

\subsection{Conductive-viscous scaling estimate}
\label{sec:scaling-rotation}

With the current prescribed, the total Lorentz torque is fixed
(Section~\ref{sec:device}). The torque on the gas is balanced by the
stresses on the walls; in the
simulations, at the current peak, more than 90\,\% of the angular momentum
that leaves the gas goes to the outer wall and 6 to 8\,\% to the end plate
(Section~\ref{sec:results-hd}).
The same
collisions that carry momentum to the walls also carry heat, and this links the
rotation to the gas heating. The scaling estimate below makes this link explicit.

Consider a column of gas rotating at $u_\theta$ in a steady state, with
uniform viscosity $\mu$ and thermal conductivity $\kappa$, the angular momentum carried to the wall by
viscosity over a length $\ell_\mu$ and the heat carried to the wall by conduction
over a length $\ell_\kappa$. The azimuthal force is balanced by viscous drag,
$|J_r|B_0\simeq\mu u_\theta/\ell_\mu^2$, so the mechanical power $|J_r|B_0u_\theta$ is
dissipated as the viscous heating $\mu u_\theta^2/\ell_\mu^2$, and the conducted
flux $\kappa\Delta T/\ell_\kappa^2$, with $\Delta T$ the temperature rise of the core
above the wall, has to carry at least this much:
\begin{equation}
u_\theta^2\lesssim\frac{\kappa}{\mu}\,\frac{\ell_\mu^2}{\ell_\kappa^2}\,\Delta T
\simeq\frac{\kappa}{\mu}\,\Delta T .
\label{eq:ceiling}
\end{equation}
The last form assumes equal lengths, $\ell_\mu=\ell_\kappa$; then the current, the
field and the conductivity drop out. Thin layers, a circulation that carries
heat and angular momentum differently, wall temperature jumps and transients
change the ratio of the two lengths, so Eq.~\eqref{eq:ceiling} is a reference
scale for $u_\theta^2/\Delta T$, not a pointwise inequality. For a
monatomic gas of atomic mass $m$, with the Chapman-Enskog ratio $\kappa/\mu=\tfrac{15}{4}\kB/m$,
this is $X\equiv mu_\theta^2/2\kB T\lesssim\tfrac{15}{8}\,\Delta T/T$, the limitation
by viscous self-heating discussed by Wijnakker and Granneman~\cite{Wijnakker1980}: in this scaling estimate the rotation stays below about
$(15/4)^{1/2}\simeq1.9$ thermal speeds $(\kB\Delta T/m)^{1/2}$ of the self-heated
gas, a rotational Mach number $u_\theta/c_s\lesssim1.5$ for $\Delta T\simeq T$.
In a mixture $m$ is the mass that carries the heat. The ratio $\kappa/\mu$ of a
mixture lies between those of its lightest and heaviest main components, since
the light species dominates the conductivity and the heavy one the viscosity,
and the light species gathers in the core and the heavy one at larger radius;
Table~\ref{tab:envelope} therefore brackets the scaling estimate with $m$ set to the
heavier (pessimistic) and the lighter (optimistic) main component.

Ohmic heating lowers $u_\theta^2/\Delta T$ below this scaling estimate. With the back-EMF,
$|J_r|=\sP(|E_r|-u_\theta B_0)$, and the momentum balance gives
$u_\theta=(|E_r|/B_0)\,f$ with $f=\Ha^2/(1+\Ha^2)$ and the Hartmann number
$\Ha^2=\sP B_0^2\ell_\mu^2/\mu$. If a fraction $\xi$ of the Ohmic heat reaches the
gas, the rest being radiated or lost to the sheaths, the heat balance gives
$u_\theta^2=(\kappa/\mu)\,\Delta T\,f/[f+\xi(1-f)]$; this factor does not exceed
one, so within the same scaling estimate Ohmic heating lowers $u_\theta^2/\Delta T$ whatever
fraction of it is radiated.

With Eq.~\eqref{eq:ceiling} for $u_{\max}$ in Eq.~\eqref{eq:S},
$\ln\alpha\lesssim S\,(\Delta T/T)\,\tfrac{15}{4}\,\Delta m/m$. For a hot core,
$\Delta T\simeq T$, and a profile factor of order unity, $S\,\Delta T/T\simeq1$,
this gives the scaling estimate
\begin{equation}
\alpha\lesssim\exp\!\left(\frac{15}{4}\,\frac{\Delta m}{m}\right),
\label{eq:envelope}
\end{equation}
which Table~\ref{tab:envelope} lists for hydrogen isotopes and for a D/T mixture
with 5\,\% of helium. For atomic H/D with the mean mass it is $\alpha\lesssim12$
(6.5 with the mass of D, 42 with that of H), against $\alpha\approx5$ measured at 1.55~kA
(Fig.~\ref{fig:alpha}); for a core at $1.5\times10^4$~K above a 300~K wall,
Eq.~\eqref{eq:ceiling} gives $u_\theta\lesssim17$~km\,s$^{-1}$, against the
13~km\,s$^{-1}$ measured in hydrogen~\cite{korobtsev1986chemical}. 
Here $S\,\Delta T/T\simeq1$ is an assumption, and the lengths $\ell_\mu$ and $\ell_\kappa$ need not be equal. The scaling estimate is also one-dimensional: it gives the radial separation only and cannot capture two-dimensional effects such as the circulation. In a device with a steady flow of gas through it, the overall separation between the head and tail streams matters more than the radial separation. In gas centrifuges an axial countercurrent multiplies the radial separation along the length of the column; the same enhancement has been studied for plasma centrifuges~\cite{Dolgolenko2023} and is a subject for future work.

\begin{table}[t]
\centering
\caption{Scaling estimate of the separation factor, $\alpha_\mathrm{est}=\exp(\tfrac{15}{4}\Delta m/m)$, of
Eq.~\eqref{eq:envelope} for fully dissociated mixtures (atomic fractions in \%), with
$S\,\Delta T/T=1$ and the heat-carrying mass $m$ set to the heavier main component
($m_\mathrm{heavy}$, pessimistic), the mean atomic mass $\bar m$, and the lighter main component
($m_\mathrm{light}$, optimistic); for D/T/He these are T and D. Atomic masses: H 1.0078, D 2.0141,
T 3.0160, $^4$He 4.0026~u.}
\label{tab:envelope}
\setlength{\tabcolsep}{3pt}
\begin{tabular}{llcccc}
\hline\hline
 & & & \multicolumn{3}{c}{$\alpha_\mathrm{est}$} \\
mixture & pair & $\Delta m$ (u) & $m_\mathrm{heavy}$ & $\bar m$ & $m_\mathrm{light}$ \\
\hline
H/D 50/50 & D/H & 1.01 & 6.5 & 12.2 & 42.3 \\
D/T 50/50 & T/D & 1.00 & 3.5 & 4.5 & 6.5 \\
D/T/He 47.5/47.5/5 & T/D & 1.00 & 3.5 & 4.3 & 6.5 \\
 & He/D & 1.99 & 11.9 & 17.8 & 40.5 \\
 & He/T & 0.99 & 3.4 & 4.2 & 6.3 \\
\hline\hline
\end{tabular}
\end{table}

\section{Results}
\label{sec:results}
\label{sec:results-hd}

The model is tested against the H/D separation experiments of
Korobtsev \textit{et al.}~\cite{korobtsev1986chemical}.

\subsection{Discharge evolution}
\begin{figure}[t]
\centering
\includegraphics[width=\columnwidth]{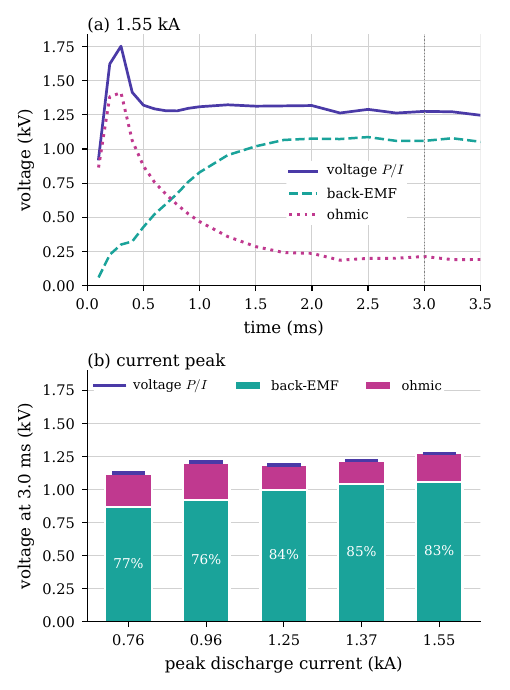}
\caption{\label{fig:voltage}Split of the discharge voltage into the back-EMF of the
rotating gas, $-(1/I)\int J_ru_{\nn\theta}B_0\,\dd V$, and the ohmic part,
$(1/I)\int\Jb\cdot\Eb'\,\dd V$, which add up to the current-weighted voltage $P/I$, with $P=\int\Jb\cdot\Eb\,\dd V$ the electrical power.
(a) Against time at 1.55~kA; dotted line: the current peak. (b) At the current
peak (3.0~ms) for the five currents, with the back-EMF fraction.}
\end{figure}

Figure~\ref{fig:voltage}(a) shows the evolution of the discharge voltage at 1.55~kA. The radial current is driven
by the neutral-frame field $E'_r=E_r+u_{\nn\theta}B_0$
(Section~\ref{sec:ohm}), so the total voltage is
the sum of an ohmic part and the back-EMF of the
rotating gas. At the start of the discharge the voltage is mostly ohmic.
As the plasma density rises and the gas spins up, the ohmic part falls and the back-EMF grows until it is the larger part.
At the current peak the back-EMF is 76 to 85\,\% of the voltage at all five currents [Fig.~\ref{fig:voltage}(b)].
Near the current peak the voltage is therefore mostly a
measure of the gas rotation, and the overestimate of $\nee$
(Section~\ref{sec:validity-plasma}), which acts only on the ohmic part, has a
limited effect on it. No measured
voltage-current characteristic of the device has been published.

The largest potential
difference is 3 to 5\,\% higher than the voltage of Fig.~\ref{fig:voltage}, the electrical power divided by the current,
because the
cathode current is imposed as a profile and the cathode tip is not an
equipotential. In the experiment the cathode is a metal electrode at a single
potential, and the current probably entered it through cathode spots. Spots are small and involve non-neutral sheaths and electrode material, so they are too
complicated to be included in the model. The model imposes a smooth current-density
profile on the cathode tip instead (Section~\ref{sec:ohm}); it is a
simplification of the discharge there.
The current distribution near the
cathode can therefore
differ from that of the experiment, whereas the total Lorentz torque, set by the
current and the magnetic field, does not; and a measured voltage would include a cathode fall that the
model does not have, of order 10 to 20~V for cathode spots~\cite{LyubimovRakhovskii1978}, small against the
kilovolt discharge voltage.

\subsection{Rotation}
\begin{figure}[t]
\centering
\includegraphics[width=\columnwidth]{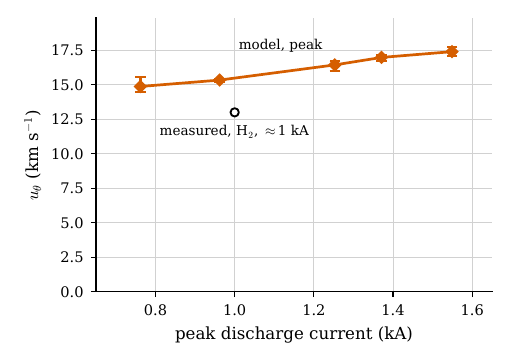}
\caption{\label{fig:rotT}Peak azimuthal speed of the gas on the midplane
against the peak current: the mean over the saved states from 2.5 to
3.5~ms, with the range in that window.
Measured: about 13~km\,s$^{-1}$ in
hydrogen at about 1~kA~\cite{korobtsev1986chemical}. The measurement is optical, a
line-of-sight average.}
\end{figure}

Figure~\ref{fig:rotT} shows the peak azimuthal speed of the gas on the midplane for the five currents, averaged over 2.5 to 3.5~ms, and the value of about 13~km\,s$^{-1}$
measured in hydrogen at 1~kA~\cite{korobtsev1986chemical}.
As an average along the line of sight (Section~\ref{sec:experiment}), the measured value is expected to lie below the peak of the gas
rotation on the midplane.

\begin{figure}[t]
\centering
\includegraphics[width=\columnwidth]{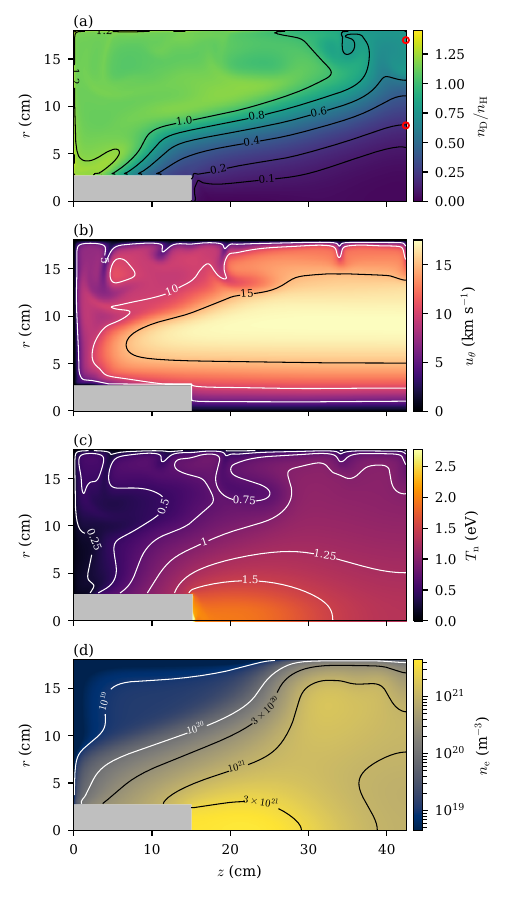}
\caption{\label{fig:maps1549}1.55~kA at $t=3.0$~ms (current peak), in the
$(z,r)$ half-domain (end plate at $z=0$, midplane at the right; gray: the
cathode), with labeled contours. (a) $n_\D/n_\mathrm{H}$, and the sampling
radii on the midplane (circles); (b) azimuthal velocity of the gas; (c) gas temperature; (d)
electron density.}
\end{figure}

Figure~\ref{fig:maps1549} shows the gas and the plasma at
1.55~kA at the current peak (3.0~ms). The gas rotates fastest in a column at
$r\approx8$ to 10~cm that extends along the whole length above the cathode [Fig.~\ref{fig:maps1549}(b)]. The outer layer, between this column and the wall
at rest, rotates more slowly and is broken up into cells by the
meridional circulation.

At the current peak the gas is close to a steady rotation: the rate of change of its angular momentum is at most 2\,\% of the Lorentz torque on the discharge.
The gas
hands the torque that it receives to the walls.
Of the angular momentum that crosses surfaces two cells inside the walls at the
current peak, 92 to 94\,\% goes to the outer wall and 6 to 8\,\% to the end plate;
the flux toward the cathode is below 0.5\,\%.

\begin{figure*}[t]
\centering
\includegraphics[width=1.5\columnwidth]{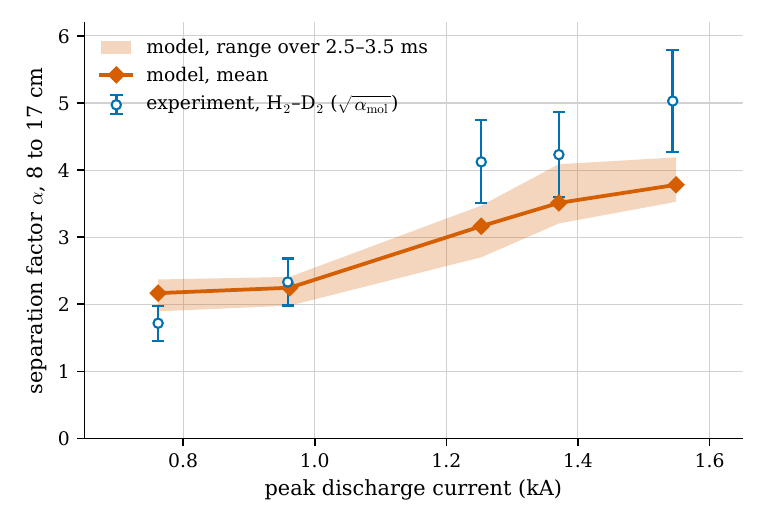}
\caption{\label{fig:alpha}Separation factor between 8 and 17~cm on the midplane
against the peak discharge current. Model: mean over the saved states from 2.5
to 3.5~ms (line) and the range from the lowest to the highest value in that
window (band). Experiment: square root of the molecular separation factor
measured in H$_2$-D$_2$, with the $\pm15$\,\% accuracy stated in
Ref.~\cite{korobtsev1986chemical}.}
\end{figure*}

\subsection{H/D separation}
\label{sec:results-separation}
The model only takes the operating conditions of the experiments of Korobtsev \textit{et al.} as inputs: the geometry of the device, the magnetic field, the pressure and the composition of the fill, and the peak current (Section~\ref{sec:device}). With these inputs it reproduces the measured separation factor
between the sampling radii,
$\alpha=(n_\D/n_\mathrm{H})_{R_2}/(n_\D/n_\mathrm{H})_{R_1}$ with $R_1=8$~cm and
$R_2=17$~cm on the midplane, to within 26\,\% at all five currents (Fig.~\ref{fig:alpha}).

In the experiments the gas was drawn into the sampler through a probe (a snorkel)
inserted into the rotating gas at the sampling radius. The probe disturbs the
flow and the composition around it in three dimensions, and the sample comes
from a finite volume around its tip; neither effect is included in the
axisymmetric model, which gives the composition of the undisturbed gas at a point.
The sampling valve of the experiment is open for 1 to 1.5~ms. We assume that it is open for 1~ms after the nominal sampling time of 2.5~ms and average the model over this window (Fig.~\ref{fig:alpha}). The sampled atoms recombine into H$_2$, HD and D$_2$ molecules. The measured atomic $\alpha$ is the square root of the separation factor of the D$_2$ and H$_2$ molecules, which is how Ref.~\cite{korobtsev1986chemical} calculates the atomic factor.
Averaging the model over a sampled volume lowers $\alpha$, but only
slightly: with $n_\D/n_\mathrm{H}$ averaged over a sphere of 1~cm radius
centered at each sampling point, the window means are 1 to
3\,\% below the point values. A valve window that opens 0.5~ms earlier, from 2.0
to 3.0~ms, gives values 1 to 10\,\% lower than those of Fig.~\ref{fig:alpha}, depending on the peak current.

The scaling estimate of Section~\ref{sec:scaling-rotation} can be compared with the runs.
Over 2.5 to 3.5~ms, the peak rotation is 86 to 97\,\% of the speed
$u_c=(\tfrac{15}{4}\kB\Delta T/m)^{1/2}$, with $m=(m_\mathrm{H}+m_\mathrm{D})/2$,
where $\Delta T$ is the rise of the maximum gas temperature on the midplane
(1.24 to 1.36~eV) above the 300~K wall. The rotation approaches this scaling estimate at the
higher currents. On the midplane at the current peak the rotation is supersonic: the
Mach number of the mixture, $u_\theta/c_s$ with $c_s=(\gamma\kB T/\bar m)^{1/2}$,
$\gamma=5/3$, the mass-averaged $u_\theta$, the number-averaged $T$ and the local
mean mass $\bar m=\rho/n$, reaches 1.3 to 1.5, close to the value of about 1.5 that
the scaling estimate gives for $\Delta T\simeq T$.
The
equilibrium separation with the computed profiles lies below the scaling estimate of Table~\ref{tab:envelope}, and
the simulated one, 2.2 to 3.8, lies further below because the outer layer is mixed
(Section~\ref{sec:scaling-mixing}).

\begin{figure}[t]
\centering
\includegraphics[width=\columnwidth]{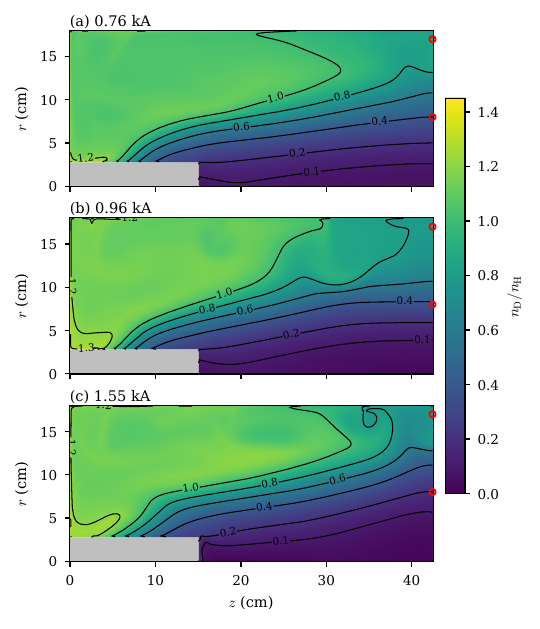}
\caption{\label{fig:dhmaps}$n_\D/n_\mathrm{H}$ at $t=3.0$~ms for peak currents
of (a) 0.76, (b) 0.96 and (c) 1.55~kA, with labeled contours (end plate at
$z=0$, midplane at the right; gray: the cathode; circles: the sampling radii,
8 and 17~cm, on the midplane).}
\end{figure}

\begin{figure}[t]
\centering
\includegraphics[width=\columnwidth]{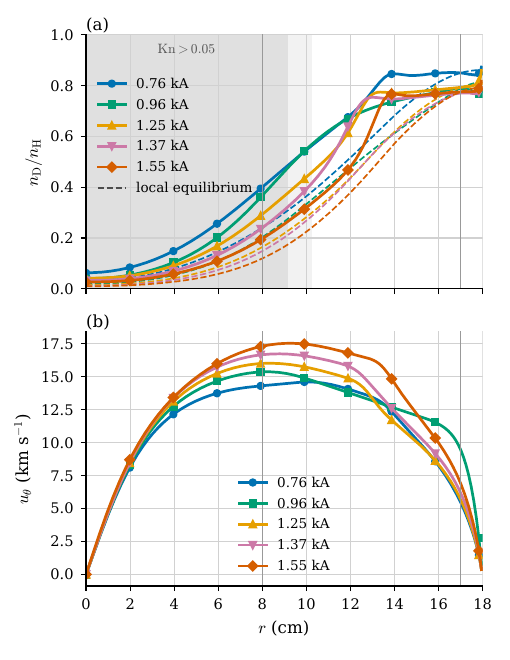}
\caption{\label{fig:profiles}Midplane profiles at $t=3.0$~ms. (a)
$n_\D/n_\mathrm{H}$ (solid) and the local equilibrium \eqref{eq:alpha} with the
simulated $u_\theta(r)$ and $\Tn(r)$, matched at 17~cm (dashed); shaded:
largest gradient Knudsen number above 0.05 (Table~\ref{tab:kn}). (b) Azimuthal speed of the gas. Vertical lines: the
sampling radii, 8 and 17~cm.}
\end{figure}

Figure~\ref{fig:dhmaps} shows where the separation takes place.
Deuterium is depleted along the axis above the cathode, where the gas is
hottest, and the depleted region widens with the current: on the midplane the contour 0.2 moves outward from panel (a) to panel (c).
Almost all of the separation comes from the depletion of deuterium at 8~cm: $n_\D/n_\mathrm{H}$ at 17~cm stays between 0.76 and 0.85, whereas at 8~cm it falls as the current rises [Fig.~\ref{fig:profiles}(a)].
Next to the end plate the outer layer is enriched in deuterium; near the midplane it is depleted in deuterium relative to the fill and nearly uniform
in composition between $r\approx14$~cm and the wall, which
limits the value at 17~cm. Next to the foot of the cathode, at the end plate,
$n_\D/n_\mathrm{H}$ is largest.
This corner is fed by the inflow along the end plate. The inflow brings gas
from the deuterium-enriched outer layer ($n_\D/n_\mathrm{H}\approx1.1$ to 1.2 at
$r\approx10$ to 18~cm) inward faster than its composition can adjust to the falling
pressure, and the gas gains deuterium on the way: along a streamline
$n_\D/n_\mathrm{H}$ rises by 13\,\% (at 1.55~kA). Two
mechanisms contribute, and their local equilibrium sets the largest rise each
can give. At 0.76 and 0.96~kA the gas cools to about 0.08~eV next to the end
plate, and thermal diffusion enriches it in deuterium (at most 19 and 18\,\%). At
1.55~kA the corner stays at 0.1 to 0.5~eV, so thermal diffusion gives little (at most
4\,\%); the gain comes mainly from pressure diffusion in the pressure rise where
the inflow slows down (at most 13\,\%).

\begin{figure}[t]
\centering
\includegraphics[width=\columnwidth]{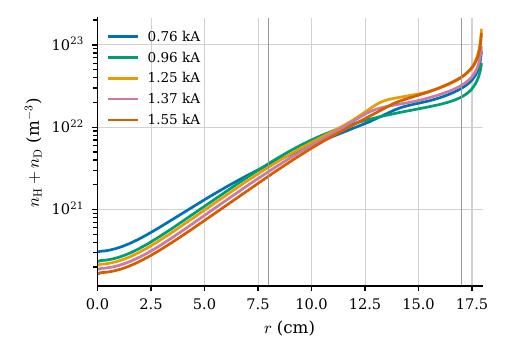}
\caption{\label{fig:profilesTn}Total neutral density on the midplane,
$n_\mathrm{H}+n_\mathrm{D}$, at $t=3.0$~ms. Vertical lines: the sampling radii,
8 and 17~cm.}
\end{figure}

Figure~\ref{fig:profiles} compares the midplane profiles at 3.0~ms with the
local equilibrium \eqref{eq:alpha}, computed from the model's own rotation and
gas temperature and matched to the model at 17~cm. Between $r\approx14$~cm and
the wall, $n_\D/n_\mathrm{H}$ varies by at most 0.05 in each run, whereas the
equilibrium keeps rising toward the wall: the outer layer is mixed by the
meridional circulation. Further in, the simulated
ratio lies above the
equilibrium, so that $\alpha$ is 60 to 64\,\% of its equilibrium
value. Figure~\ref{fig:profilesTn} shows the neutral density on the midplane.

\subsection{Mixing against separation}
\label{sec:scaling-mixing}

\begin{figure}[t]
\centering
\includegraphics[width=\columnwidth]{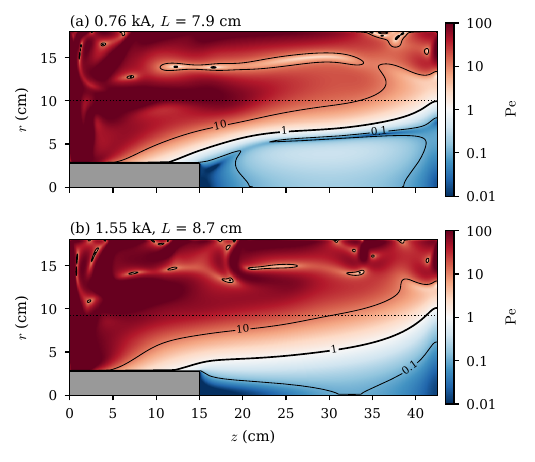}
\caption{\label{fig:peclet}P\'eclet number of the meridional circulation,
$\mathrm{Pe}=|\ub_p|L/D_\mathrm{HD}$ [Eq.~\eqref{eq:peclet}], at the current peak ($t=3.0$~ms) for (a) 0.76 and
(b) 1.55~kA, in the $(z,r)$ half-domain (end plate at $z=0$, midplane at the right;
gray: the cathode). Red: the circulation mixes faster than the species diffuse;
black lines: $\mathrm{Pe}=0.1$, 1 (thick) and 10; dotted: the rotation peak on the midplane.}
\end{figure}

Separation requires radial diffusion of the species against their mutual
friction (Section~\ref{sec:neutral}), whereas the meridional circulation driven by
the end plate carries the species through the $(r,z)$ plane. The
competition is measured by a P\'eclet number of the circulation, the ratio of the
radial diffusion time $\tau_D=L^2/D_\mathrm{HD}$ across the layer to the turnover time $\tau_c=L/|\ub_p|$ of the circulation,
\begin{equation}
\mathrm{Pe}=\frac{\tau_D}{\tau_c}=\frac{|\ub_p|\,L}{D_\mathrm{HD}},
\label{eq:peclet}
\end{equation}
where $\ub_p$ is the
meridional gas velocity, $L$ the width of the outer layer (from the rotation peak to
the wall) and $D_\mathrm{HD}$ the binary diffusion coefficient in the first
Chapman-Enskog approximation with the collision integrals of
Section~\ref{sec:neutral}. Figure~\ref{fig:peclet}
maps it at the current peak. On the midplane $\mathrm{Pe}$ is below one at the inner
sampling radius (0.09 to 0.25 at 8~cm), about 3 at 12~cm and 100 to 190 at 17~cm: in
the outer layer advection is much faster than diffusion on the scale $L$, whereas
inside the $\mathrm{Pe}=1$ contour the composition can relax toward the local
equilibrium of Section~\ref{sec:scaling-alpha}. Inside $r\approx10$~cm $D_\mathrm{HD}$ is
the continuum value in a rarefied gas (Section~\ref{sec:validity-neutral}). A large
Pe alone does not imply that a steady circulation homogenizes the composition
across its streamlines, and the centrifugal drift also sets a time scale; the
P\'eclet number is used here as an interpretive diagnostic. The simulations show a
nearly uniform composition in the outer layer, consistent with mixing by the
circulation, although an isotope-flux budget would be needed to establish the
cause: on the midplane $n_\D/n_\mathrm{H}$ varies by at most 0.05 between
$r\approx14$~cm and the wall in each run, whereas the equilibrium keeps rising toward the
wall, and $\alpha$ between 8 and 17~cm is 60 to 64\,\% of its equilibrium value
(Section~\ref{sec:results-separation}).

\section{Discussion: limits of validity}
\label{sec:validity}

\subsection{Continuum neutral description}
\label{sec:validity-neutral}

The discharge heats the core gas on the midplane to $\Tn\approx1.2$ to 1.4~eV.
The rotation expels the gas toward the wall, and the number density on the
axis falls 65 to 115 times below the fill. Whether the continuum description holds
depends on the mean free path relative to the physical gradient lengths. Table~\ref{tab:kn} lists, on the midplane at the current peak,
the mean free path $\lambda=(\mu/p)(\pi\kB T/2\bar m)^{1/2}$ computed from the
viscosity $\mu$ of the model~\cite{ChapmanCowling1970}, which uses the temperature-dependent
collision integrals of Section~\ref{sec:neutral}, the pressure $p$ and the local mean atomic mass $\bar m$ of the gas, and the gradient-length Knudsen
numbers~\cite{Boyd1995} $\Kn_n=\lambda|\partial_r\ln n|$ of the density and
$\Kn_x=\lambda|\partial_r\ln x_\D|$ of the composition. Those of the temperature,
$\lambda|\partial_r\ln T|$, and of the shear rate,
$\lambda|\partial_ru_\theta-u_\theta/r|/\max(u_\theta,c_s)$, are smaller at these radii (at most 0.05 and 0.26). On the axis, a symmetry extremum, the gradient
measures vanish and are not informative.

\begin{table*}[t]
\centering
\caption{Rarefaction on the midplane of the H/D simulations at the current peak,
$t=3.0$~ms. $\lambda$: viscosity-based mean free path with the transport of the
model; $\Kn_n$, $\Kn_x$: gradient-length Knudsen numbers of the density and of
the D mole fraction.}
\label{tab:kn}
\setlength{\tabcolsep}{3pt}
\resizebox{\ifdim\width>\linewidth\linewidth\else\width\fi}{!}{%
\begin{tabular}{l|*{5}{c}|*{5}{c}|*{5}{c}}
\hline\hline
 & \multicolumn{5}{c|}{$\lambda$ (mm)} & \multicolumn{5}{c|}{$\Kn_n$} & \multicolumn{5}{c}{$\Kn_x$} \\
\cline{2-6}\cline{7-11}\cline{12-16}
$I_\mathrm{peak}$ (kA) & 0.76 & 0.96 & 1.25 & 1.37 & 1.55 & 0.76 & 0.96 & 1.25 & 1.37 & 1.55 & 0.76 & 0.96 & 1.25 & 1.37 & 1.55 \\
\hline
$r$ (cm) & & & & & & & & & & & & & & & \\
2  & 18.9 & 23.3 & 26.3 & 29.6 & 33.5 & 0.59 & 0.74 & 0.83 & 0.89 & 1.00 & 0.45 & 0.64 & 0.57 & 0.62 & 0.74 \\
4  & 9.7 & 11.6 & 13.2 & 15.2 & 17.7 & 0.36 & 0.48 & 0.55 & 0.63 & 0.71 & 0.25 & 0.36 & 0.38 & 0.46 & 0.54 \\
8  & 2.6 & 2.4 & 2.7 & 3.0 & 3.5 & 0.075 & 0.089 & 0.10 & 0.12 & 0.14 & 0.034 & 0.045 & 0.049 & 0.065 & 0.078 \\
12 & 0.90 & 0.79 & 0.64 & 0.68 & 0.79 & 0.021 & 0.014 & 0.024 & 0.025 & 0.024 & 0.006 & 0.003 & 0.009 & 0.011 & 0.012 \\
17 & 0.22 & 0.35 & 0.16 & 0.22 & 0.17 & 0.008 & 0.011 & 0.007 & 0.008 & 0.006 & $<$0.001 & $<$0.001 & $<$0.001 & $<$0.001 & $<$0.001 \\
\hline\hline
\end{tabular}}
\end{table*}

\paragraph{Rarefied core.}
The largest gradient Knudsen number exceeds 0.05 inside $r=9.1$ to 10.3~cm,
the radius being larger at the higher currents, and reaches 0.4 to 1.0 at 2 to 4~cm. A value of
about 0.05 is used as an approximate indicator of continuum breakdown in
hypersonic flows~\cite{Boyd1995}; it is not a sharp validity boundary here, but
values of order 0.1 to 1 mean that the Navier-Stokes stress, the Fourier heat flux
and the local inter-species friction are no longer justified. The inner sampling
radius, $R_1=8$~cm, lies where $\Kn_n=0.08$ to 0.14, at the edge of this region,
while the outer sample ($R_2=17$~cm, $\Kn_n\approx0.01$) lies in the collisional
region. The composition at 8~cm, and therefore the separation factor between $R_1$ and $R_2$, is computed
where the continuum closure is marginal. Kinetic effects can change the
transport, the response of the boundaries and the relaxation of the composition;
their sign has not been established here.

The mixture corrections of the transport coefficients do not remove this
limitation. Wilke's rule~\cite{Wilke1950,MasonSaxena1958} apportions the viscosity and conductivity of a collision-dominated
mixture between the species but does not extend the closure to rarefied
conditions; the empirical Knudsen factor $(1+\Kn_s)^{-1}$ uses the vessel radius as the
length scale, so at 2 to 4~cm on the midplane, where the gradient Knudsen numbers
are 0.4 to 1.0, it lowers $\mu_s$ and $\kappa_s$ by only 6 to 19\,\% for H and
6 to 22\,\% for D; and the inter-species friction, which sets
the rate of separation, keeps its continuum form.

This region is also where the plasma forces on the gas are damped
(Section~\ref{sec:numerics}). The ionization fraction is high there, and Ohm's
law, which drops the inertia of the plasma and lets the gas take up the
Lorentz force, is at its limit; the gas of the core therefore receives only
part of the Lorentz torque that acts on it.

\paragraph{Composition near the axis.}
We evaluated the local equilibrium \eqref{eq:alpha} with the computed
$u_{\nn\theta}(r)$ and $\Tn(r)$. On the midplane, over the window 2.5 to 3.5~ms, the
simulated $\alpha$ between 8 and 17~cm is 62, 64, 61, 60 and 60\,\% of its
equilibrium value at 0.76, 0.96, 1.25, 1.37 and 1.55~kA, respectively, because the gas at 8~cm is less depleted of deuterium than the
equilibrium with the local rotation would make it (Fig.~\ref{fig:profiles}).
The separation is therefore limited both by the computed rotation and gas
temperature, which rely on the continuum closures in the rarefied core, and by
the mixing of the meridional circulation (Section~\ref{sec:results-hd}). A
20\,\% error in $\Tn$ produces a comparable error in the equilibrium
$\ln\alpha$.

\paragraph{Numerical accuracy where the closure holds.}
Outside this region the accuracy of the computed rotation depends on how well
the solver represents the layer on the end plate and the circulation it
drives. Section~S1 of the supplementary material measures this in tests of the gas
solver alone, including one at the Mach number and the resolution of that
layer in the simulations. Those tests do not reach the strongly non-uniform
viscosity of the simulations or the coupling to the plasma.

\subsection{Plasma density}
\label{sec:validity-plasma}

A further limitation concerns the plasma density.
In the model, \eqref{eq:ne} contains no ion advection, so
plasma is redistributed only by ambipolar diffusion, which is slow across the
field, and lost by volume recombination and at the walls. The ion current would carry the plasma to the cathode, where the ions are neutralized. The model has neither this advection nor an ion sink at the cathode (Section~\ref{sec:plasma}), so the plasma accumulates and the density is overestimated.
Forced quasineutrality also removes the
non-neutral layers at the electrodes, which in reality affect the plasma supply
and carry part of the voltage. Averaged over the gas volume at 2.5~ms, the simulated density at 0.96~kA is $1.2\times10^{20}$~m$^{-3}$, 6 to 12 times the $(1$ to $2)\times10^{19}$~m$^{-3}$ measured at 1~kA~\cite{korobtsev1986chemical}.

Early in the pulse the density is higher still. In a column on the axis above
the cathode tip, where the gas is almost fully ionized, $\nee$ at the four higher currents exceeds the upper density limit of the rate-coefficient
fits, $10^{22}$~m$^{-3}$ (Section~\ref{sec:atomic}), between 0.3 and 1.4~ms,
in at most 4\,\% of the cells. The highest density reached is $4.3\times10^{22}$~m$^{-3}$, at 1.55~kA. However, in the window in which the separation is evaluated the largest density is inside the range of the fits.

The Lorentz torque on the discharge is fixed by the prescribed current
and does not depend on $\nee$.
A density that is too high acts instead through the conductivity: a
larger $\sP$ lowers the neutral-frame field $E'_r$ needed to carry the current,
and with it the Joule dissipation $J_r^2/\sP$. It also raises the ionization and
radiation losses of the electrons, and it strengthens the plasma-neutral
coupling in regions where the real plasma would be depleted. At fixed current these effects act in
opposite directions on the gas heating, so the computed gas temperature, and
hence the separation, carries an uncertainty from $\nee$ whose sign has not been
determined.

\subsection{Molecular processes}
\label{sec:validity-molecules}

Another limitation is the representation of the hydrogen isotopes by atoms
only (Section~\ref{sec:config}).
The model contains neither the
molecules nor the electron-molecule collisions, including those of
vibrationally excited molecules, from which dissociation and dissociative
attachment proceed at much lower electron energies, and this affects three of its
results.

First, the plasma density. Electron-impact dissociation, dissociative
ionization and recombination assisted by molecules (through H$_2^+$, HD$^+$,
D$_2^+$ and negative ions) are electron energy losses and electron sinks that
the atomic model does not have; molecular ionization channels also add ion
production. Their net effect on $\nee$ requires a model with molecules.

Second, the gas temperature. At fixed current a lower $\nee$ would mean a lower
conductivity, a larger neutral-frame field and a larger Joule dissipation
(Section~\ref{sec:validity-plasma}), and the dissociation of a molecule releases
atoms with kinetic energies of a few eV (Franck-Condon), which heat the gas;
the energy spent on dissociation and vibrational excitation, on the other hand,
is taken from the electrons. The net change of the gas temperature is therefore
not determined without a molecular model.

Third, the composition of the neutral gas near the walls. Recombination on the
outer wall ($r=r_{\max}$) and on the end plate ($z=0$) makes the gas next to
them partly molecular, and pairs the isotopes into H$_2$, HD and D$_2$. The atomic
and molecular fractions differ there from those of the atomic model, including
at the outer sampling radius, 17~cm, only 1~cm from the wall. The total isotope
abundance, which the sampler infers, is a different quantity from the atomic
fractions, so molecule formation alone does not invalidate the computed H/D
ratio; it changes the transport of the isotopes near the walls.
A model with
molecular species, wall recombination and electron-molecule collisions is
being developed.

\subsection{Centrifugal stability of the outer layer}
\label{sec:scaling-stability}

In the simulations the gas rotates fastest in a column at
$r\approx8$ to 10~cm, and the outer layer, between this column and the wall at
rest, contains axisymmetric circulation cells (Section~\ref{sec:results-hd}).
A local diagnostic indicates whether this layer is susceptible to centrifugal
instability. In an inviscid rotating flow, axisymmetric displacements are
unstable where the angular momentum decreases outward, that is, where the
squared epicyclic frequency $\kappa^2=r^{-3}\,\dd(ru_\theta)^2/\dd r$ is
negative~\cite{Rayleigh1917,Chandrasekhar1961}; this is the case in part of the region between the
rotation peak and the wall. The radial stratification in the centrifugal
acceleration $g=u_\theta^2/r$ opposes the displacement and is measured by the
squared buoyancy frequency
\begin{equation}
N^2=g\left(\frac{\partial\ln\rho}{\partial r}-\frac{1}{\gamma}\frac{\partial\ln p}{\partial r}\right),
\label{eq:N2}
\end{equation}
with $\rho$ and $p$ the mass density and the pressure of the gas; it is positive here, where the gas is compressed and cooled toward the wall. We use
the radial part of the Solberg-H{\o}iland condition~\cite{Tassoul1978},
$\kappa^2+N^2<0$, as a local diagnostic; it is not a stability analysis of the
two-dimensional, viscous, diffusive and stratified mixture, and it gives neither
growth rates nor mode structures. Between the rotation peak and the wall
$\kappa^2<0$ in about a third of the whole gas volume (about half its mass), and
$\kappa^2+N^2<0$ in 20 to 23\,\% of the volume (40 to 44\,\% of the mass), mostly in
the outer part of the layer ($N^2/|\kappa^2|=0.3$ to 0.4 at 17~cm; at 12~cm
it exceeds one in four of the five runs). As an analogy only, the Taylor-Couette Reynolds number
$\mathrm{Re}_\mathrm{TC}=u_{\max}\,d/\nu$, with $d$ the gap between the rotation peak
and the wall and $\nu$ the kinematic viscosity of the layer, is 400 to 900,
far above the onset of Taylor vortices between cylinders with the outer one at
rest (critical value $\mathrm{Re}_c\approx68$ for a radius ratio of 0.5~\cite{Taylor1923,Chandrasekhar1961});
the present flow has no inner cylinder, is compressible, stratified, body-forced
and crossed by the circulation, so this threshold does not transfer. The
axisymmetric model can represent axisymmetric centrifugal cells, and the cells
of the simulations may have several causes; it cannot represent
non-axisymmetric modes or their nonlinear transport. The mixing of the outer
layer, and with it the composition at the outer sampling radius, is therefore
uncertain.

\section{Conclusions}
\label{sec:conclusions}

We have developed a time-dependent, axisymmetric multifluid model of a partially
ionized plasma centrifuge, in which the properties of the plasma and the gas, such as the temperatures and ionization
fraction, are computed rather than imposed from measurements or adjusted as free
parameters, and applied it to the H/D separation experiments of
Korobtsev \textit{et al.}~\cite{korobtsev1986chemical} at 0.3~Torr and 0.5~T for
peak currents of 0.76 to 1.55~kA.
The model reproduces the measured separation factors to within 26\,\% at all five discharge currents (Fig.~\ref{fig:alpha}). At the current peak 76 to 85\,\% of the computed voltage is the back-EMF of the rotating gas;
no measured voltage-current curve is available for comparison. At 0.96~kA the gas rotates at a peak of 15.3~km\,s$^{-1}$ on the midplane, slightly above the value of about 13~km\,s$^{-1}$ measured in hydrogen at about 1~kA, as expected because the measurement is a line-of-sight average.

In the simulations two features set the separation. The peak rotation is
86 to 97\,\% of a conductive-viscous scaling estimate evaluated with the computed
temperature. The outer part of
the layer between the rotation peak and the wall is nearly uniform in
composition, with a
P\'eclet number that rises from about 3 at 12~cm to 100 to 190 near the wall, so the
separation comes mainly from the depletion of
deuterium at the inner sampling radius.

The main limitations are the continuum description of the rarefied core, the closures of the
plasma density and of the electrode layers, and the omission of molecules. The
gradient Knudsen number exceeds 0.05 inside $r\approx9$ to 10~cm, which includes the
inner sampling radius. At 0.96~kA the
simulated mean density is 6 to 12 times the
measured one, because the density equation contains no ion advection: the ion current does not carry the plasma to the cathode, and the plasma accumulates.
Molecules and their reactions are absent, with effects on the
plasma density and the gas temperature whose sign is not established.
The local stability diagnostic indicates that the outer
layer is susceptible to centrifugal instability, and non-axisymmetric motions are
outside the model.

These limits point to the next steps: a kinetic description of the rarefied core; the ion transport
that carries the cross-field current; molecular species with wall recombination and electron-molecule
chemistry; and
three-dimensional simulations of the outer layer.
The aim beyond them is a model with more species and a steady flow of gas through the device, which can treat mixtures of deuterium, tritium and helium in continuous operation: the mass-selective pumping in the fusion fuel cycle that motivates the plasma centrifuge.

\begin{acknowledgments}
This work was supported by Marathon Fusion. This material is based upon work
supported by the U.S. Department of Energy, Advanced Research Projects
Agency-Energy (ARPA-E), under Award No.~\mbox{DE-AR0002075}. We thank Ian Wojtowicz
and Conner Fransoo for their work on an earlier steady-state magnetohydrodynamic
model, based on the framework of Wijnakker \textit{et al.}~\cite{Wijnakker1979},
which helped our understanding of the plasma centrifuge.

During the preparation of this manuscript, the authors used AI tools to
support the development of the simulation code and the analysis scripts, the checks
of the manuscript against the code and the data, and the drafting and editing of the
text. The authors reviewed and edited the content and take full responsibility for it.
\end{acknowledgments}

\section*{Data availability statement}
The data that support the findings of this study are available upon
reasonable request from the authors.

\ifdefined\submissionlayout\else\section*{References}\fi
\bibliography{references}

\end{document}


\title{Supplementary material for: A multifluid model of hydrogen isotope separation in a partially ionized plasma centrifuge}

\author{S. Cao}
\author{M. D. Acciarri}
\affiliation{Marathon Fusion, San Francisco, CA, USA}

\maketitle

Sections, equations, figures and tables numbered without the prefix S refer to the main text.

\section{Verification of the neutral-gas solver}
\label{app:neutral-tests}

The rotation that the gas reaches is set by the balance between the azimuthal
plasma force and the angular momentum carried to the walls. Across the vessel
most of this angular momentum is carried by the meridional circulation;
viscosity hands it to the walls only in the thin layers next to them. The end
plate drives such a circulation. Being at rest, it brakes the gas next to it;
the pressure gradient that balances the centrifugal force in the bulk then
drives this gas inward along the plate, and the return flow carries angular
momentum from the bulk toward the walls. How much of it ends on the plates and how much on the side
wall depends on the aspect ratio of the vessel: in the long vessel of the main
text the side wall takes most of it (Section~VI), whereas the
plates are expected to take a larger share in a short vessel. The same circulation mixes the
composition that the centrifugal force separates. The computed rotation, voltage and separation factor
therefore depend on how accurately the gas solver represents a layer whose
thickness scale, $\delta=\sqrt{\nu/\Omega}$, with $\nu$ the kinematic viscosity of the gas and $\Omega$ its angular velocity, is a few cells of the grid. This
section tests that part of the model alone. Table~\ref{tab:verif} summarizes the results.

\begin{table}[t]
\centering
\caption{Verification of the gas solver without plasma: headline results
of the tests of this section. Exact: difference from the exact
solution; OpenFOAM: difference from OpenFOAM~13 under identical conditions.
Cells are counted along the radius; $U$ is the driving speed.\label{tab:verif}}
\setlength{\tabcolsep}{2pt}\footnotesize
\resizebox{\columnwidth}{!}{%
\begin{tabular}{lll}
\hline\hline
Test & Ref. & Result \\
\hline
Decay of a rotation, 40 cells & exact & $-0.084\,\%$ in rate \\
Steady Couette profile, 40 cells & exact & $1.6\times10^{-4}\,U$ \\
Viscous heat / kinetic energy removed & exact & 1.0000 \\
Sealed cylinder, low $M$: mass / ang.\ mom. & exact & $3\times10^{-16}$ / $6\times10^{-16}$ \\
Separation in rigid rotation, 40 cells & exact & $0.00\,\%$ in $\ln\alpha$ \\
Thermal-diffusion equilibrium & exact & $4\times10^{-4}$ in $\ln(x_\mathrm{D}/x_\mathrm{H})$ \\
Spin-down: $T_{1/4}$ / core rotation & OpenFOAM & $-0.6$ to $-0.7\,\%$ / $\le0.9\,\%$ \\
Spin-down: peak inflow of the plate layer & OpenFOAM & $-0.8$ to $-2.4\,\%$ \\
Plate layer, rotating core: inflow / $u_\theta$ & OpenFOAM & $+0.5\,\%$ / $-0.4\,\%$ \\
Plate layer, rotating core: pumping & OpenFOAM & $+2.9$ to $+6.3\,\%$ \\
Plate layer, hot core: inflow / $u_\theta$ & OpenFOAM & $+3.9\,\%$ / $-3.0\,\%$ \\
Plate layer, hot core: pumping / $T$ & OpenFOAM & $+2.4$ to $+3.4\,\%$ / $\le0.3\,\%$ \\
\hline\hline
\end{tabular}}
\end{table}

\subsection{Setup, dimensionless numbers and references}
\label{app:neutral-tests-setup}

All tests use the gas solver of Section~IV of the main text without plasma, in
a plain cylinder of radius $R=\SI{5}{\centi\metre}$ and half-length $h$, with
a plate at rest at $z=0$, a side wall at rest at $r=R$ and a symmetry plane at
$z=h$. The scheme, its parameters, the wall conditions and the transport
coefficients are those of the simulations of this paper: second-order
(limited) reconstruction, the low-Mach scaling of the flux dissipation, with
both dissipative terms of the low-Mach flux (the mass flux on pressure jumps and
the damping of jumps of the normal velocity) acting on the jumps between the
reconstructed face values, and no stabilizing term ($\epsilon=0$ in
Eq.~(43); Section~IV).
The rotating tests use the resolution of the simulations next to the end
plate, 2.6 cells per $\delta$ along $z$ and 3 along $r$. Unless stated otherwise, the gas is atomic
hydrogen at \SI{300}{\kelvin} with walls at \SI{300}{\kelvin}, carried as two
identical fluids so that the two-fluid coupling is exercised. Three numbers
define a rotating test: the Reynolds number $Re=\Omega R^{2}/\nu$, set through
the pressure; the Mach number $M=\Omega R/c_s$ of the rotation at the wall
radius, with $c_s$ the sound speed at the wall temperature; and the number of
cells per $\delta$. The rotating tests use $\Omega\,\Delta t=0.01$; in the simulations the gas step of
\SI{0.25}{\micro\second} gives $\Omega\,\Delta t\approx0.05$--0.06 at the radius
of fastest rotation.

Two kinds of reference are used. (i) Exact solutions, for the tests of
single operators and of conservation. (ii) For the rotating flows, which have
no closed solution in a finite cylinder, the same problem solved by an
established, independently developed code, OpenFOAM~13 (OpenFOAM Foundation
release~13)~\cite{Weller1998}, with its compressible pressure-based solver
(\texttt{foamRun} with the \texttt{fluid} module and the \texttt{viscousHeating} fvModel) on an axisymmetric wedge grid with
swirl, for a single gas:
the same geometry, gas (mass, heat capacity and the viscosity and
conductivity tables of the solver), initial state, forces and walls, and a
grid refined until its own result changes by less than \SI{1}{\percent}. These
comparisons are made for a single gas (in the gas solver, two identical fluids), so
they test the operators and the exchange machinery but not the transport of
unequal-mass species, which is tested against exact solutions only (separation and
thermal diffusion below). The OpenFOAM case generators, meshes and settings are part
of the verification suite of the code and are available from the authors upon request.
Both
codes are run with no-slip walls at a fixed temperature for these
comparisons; wall slip and the temperature jump are tested against exact
solutions. The energy equation of OpenFOAM's solver includes the viscous work
and the work of body forces only when they are added as source terms; both are
added here; without them the kinetic energy lost to friction is not
dissipated where it is lost.

\subsection{Single operators and conservation}
\label{app:neutral-tests-operators}

\begin{figure*}[tp]
\includegraphics[width=\textwidth]{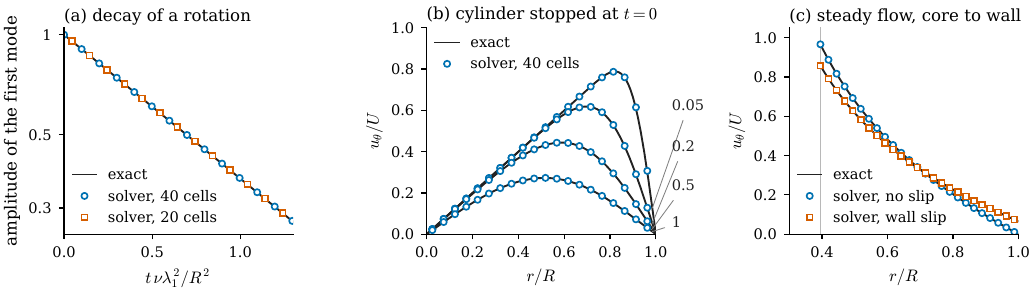}
\caption{\label{fig:neutral-tests-operators}Tests of the azimuthal viscous
operator and the walls, without meridional flow. (a) Amplitude of the first
radial mode of a rotation that decays in a cylinder at rest. (b) Rigid
rotation in a cylinder that is stopped at $t=0$: profiles at
$t\,\nu\lambda_1^{2}/R^{2}=0.05$, 0.2, 0.5 and 1. (c) Steady flow between a
core held in rigid rotation ($r<0.4R$, left of the gray line) and the wall at
rest, without slip and with wall slip at a Knudsen number of 0.07.}
\end{figure*}

\emph{Decay of a rotation.} With no friction on the plate and no meridional
flow, a rotation $u_\theta\propto J_1(\lambda_1r/R)$ in a cylinder at rest
keeps its shape and decays at the rate $\nu\lambda_1^{2}/R^{2}$, where
$\lambda_1=3.830$ is the first root of $J_1(\lambda)+(L_\mathrm{s}/R)\lambda
J_1'(\lambda)=0$ and $L_\mathrm{s}$ the slip length of the wall (3.832 without
slip)~\cite{Batchelor1967}. The test has a peak Mach number of 0.05 and a time
step of $6\times10^{-4}$ of the decay time. It involves the implicit
azimuthal viscous solve, its time integration and the outer wall. The
measured rate is within 0.16, 0.084 and \SI{0.063}{\percent} of the exact
one on 20, 40 and 80 cells
(Fig.~\ref{fig:neutral-tests-operators}a, Table~\ref{tab:neutral-tests-operators}).
Raising the time step from $1.5\times10^{-4}$ to $9.5\times10^{-3}$ of the
decay time changes the rate by \SI{0.5}{\percent}. For a cylinder stopped at $t=0$
(Fig.~\ref{fig:neutral-tests-operators}b) the profile differs from the Bessel
series by $7.5\times10^{-3}$, $1.8\times10^{-3}$, $7.5\times10^{-4}$ and
$4\times10^{-4}$ of the initial wall speed $U$ at the four times shown, the
largest difference being next to the wall, where the initial profile is
discontinuous.

\emph{Steady profile and wall slip.} A core of gas, $r<0.4R$, is held in rigid
rotation by an azimuthal force, and the gas between it and the wall at rest
takes the profile $u_\theta=Ar+B/r$ (Fig.~\ref{fig:neutral-tests-operators}c).
The solver reproduces it to $1.6\times10^{-4}$ and $8.3\times10^{-5}\,U$ on
40 and 80 cells, with $U$ the speed of the core edge. On 20 cells the
error is $6.7\times10^{-2}\,U$: there the low-Mach scaling of the flux
dissipation lets a spurious meridional circulation of $0.09\,U$ grow in this
purely azimuthal flow (Mach number 0.1). The simulations are not run at that
resolution. At a Knudsen number of 0.07 (mean free path over gap width) the profile
follows the slip condition that the solver applies,
$u_\mathrm{s}=-L_\mathrm{s}\,\partial_ru_\theta$, to $3.4\times10^{-4}$,
$7.4\times10^{-5}$ and $3.1\times10^{-5}\,U$ on 20, 40 and 80 cells for a
momentum accommodation coefficient $\sigma_v=1$, and to $3.7\times10^{-4}$,
$8.4\times10^{-5}$ and $3.4\times10^{-5}\,U$ for $\sigma_v=0.5$.

\begin{table}[b]
\centering
\caption{Tests of single operators and of conservation, with the scheme of the
simulations. Rates and temperatures: relative difference from the exact
value. Profiles: largest difference in units of the driving speed $U$.
Variable viscosity: difference of the azimuthal force from the exact one,
relative to its maximum. Separation: difference of the steady $\ln\alpha$ from
the exact value and of its build-up time from the diffusion solution with
$D_\mathrm{CE}$. Thermal diffusion: largest difference in
$\ln(x_\mathrm{D}/x_\mathrm{H})$ from the equilibrium, axis at 3000 and
\SI{12000}{\kelvin}. Cells are counted along the radius.}
\label{tab:neutral-tests-operators}
\setlength{\tabcolsep}{2pt}\footnotesize
\begin{tabular}{ll}
\hline\hline
Test & Result \\
\hline
Decay rate, 20/40/80 cells & $-0.16$/$-0.084$/$-0.063\,\%$ \\
Steady profile, 20/40/80 cells & $670^{\,a}$/$1.6$/$0.83\times10^{-4}$ \\
Stopped cylinder, 4 times & $75$--$4\times10^{-4}$ \\
Wall slip, $\sigma_v=1$ / 0.5, 40 cells & $7.4$ / $8.4\times10^{-5}$ \\
Viscous heat / kinetic energy removed & 1.0000 \\
Peak temperature rise & $+3.9\,\%$ \\
Variable viscosity, 20/40/80 & $15$/$3.7$/$1.6\times10^{-4}$ \\
Sealed cylinder, low $M$ / shocks: & \\
\quad mass of each fluid & $3\times10^{-16}$ / $3\times10^{-16}$ \\
\quad angular momentum & $6\times10^{-16}$ / $5\times10^{-5}$ \\
\quad total energy & $5\times10^{-5}$ / $3\times10^{-7}$ \\
Separation, 80 / 40 / 20 cells: & \\
\quad steady $\ln\alpha$ & $0.00$ / $0.00$ / $-0.49\,\%$ \\
\quad build-up time & $+1.4$ / $+0.6$ / $-2.7\,\%$ \\
Thermal diffusion, 3 / 12~kK & $4$ / $4\times10^{-4}$ \\
\hline\hline
\end{tabular}

\raggedright\footnotesize $^a$ Spurious meridional flow at this resolution (see the text).

\end{table}

\emph{Viscous heating.} The same flow at a core speed of $0.5\,c_s$, with
adiabatic walls, the core held at \SI{300}{\kelvin} and one cell along $z$, has a closed-form temperature profile
if the viscosity and the conductivity are uniform. The heat deposited equals
the kinetic energy that the viscous solves remove, to four digits. The peak
temperature rise is \SI{3.9}{\percent} above the closed form at
$\Delta t=\SI{1}{\micro\second}$ (\SI{25.5}{\kelvin} against
\SI{24.5}{\kelvin}), with the viscosity varying by \SI{5}{\percent} across
the gap.

\emph{Variable viscosity.} The terms of the viscous force that contain the
gradient of the viscosity are tested in three ways. For smooth velocity fields
and a viscosity that varies threefold across the cylinder, the azimuthal
force applied by one step differs from the exact
$r^{-2}\partial_r[r^{2}\mu(\partial_ru_\theta-u_\theta/r)]+\partial_z(\mu\,\partial_zu_\theta)$
by $1.5\times10^{-3}$, $3.7\times10^{-4}$ and $1.6\times10^{-4}$ of its
maximum on 20, 40 and 80 cells. In the heated flow of the previous paragraph, where the viscosity
is \SI{5}{\percent} higher in the gap than at the core, the velocity
agrees to $1.7\times10^{-4}\,U$ with the solution of
$\mu r^{3}\partial_r(u_\theta/r)=\mathrm{const}$ for the viscosity of the
solver, and differs by $10^{-2}\,U$ from the profile for uniform viscosity. With a
viscosity that changes by \SI{30}{\percent} from cell to cell in a band and a
time step 2000 times the explicit limit, the kinetic energy never grows.

\emph{Conservation.} A sealed cylinder with the viscosity and the conductivity
reduced by $10^{8}$ is started from a swirling flow that is not an
equilibrium and followed for three turnovers, at a Mach number of 0.07 and in
a violent case in which shocks form and the Mach number of the meridional
flow exceeds 2. With the scheme of the simulations the mass of each fluid is
conserved to round-off in both cases on $40\times40$ cells, and so is the
angular momentum at low Mach number. With shocks the angular momentum changes
by $5\times10^{-5}$ in the advection step. The total energy is
conserved to $3\times10^{-7}$ with shocks and to $5\times10^{-5}$ at low Mach
number, where the low-Mach scaling of the flux dissipation sets the error.

\emph{Separation in rigid rotation.} A 50/50 H--D mixture at \SI{1}{Torr}
and \SI{300}{\kelvin} rotates rigidly at $M=0.5$ (mean mass) in a short
cylinder whose side wall co-rotates and whose plate exerts no azimuthal
friction, so that rigid rotation is an exact steady state without meridional
flow. Started from a uniform composition, each species relaxes by radial
diffusion against the H--D friction to its own Boltzmann distribution,
$\ln(n_\mathrm{D}/n_\mathrm{H})-\ln(n_\mathrm{D}/n_\mathrm{H})_{r=0}=(m_\mathrm{D}-m_\mathrm{H})\Omega^{2}r^{2}/2\kB T$.
With the friction of the simulations the steady separation factor between
$0.3R$ and $0.9R$ is exact to \SI{0.3}{\percent} in $\ln\alpha$ against the
isothermal value on 40 and 80 cells
(0.00, 0.00 and
$-0.49\,\%$ against the value at the temperature of the gas on 80, 40 and 20 cells; the small difference from
the isothermal value is the heating by the initial adjustment), and the time
in which the separation builds up agrees with a one-dimensional diffusion
solution with the first Chapman--Enskog interdiffusion
coefficient~\cite{ChapmanCowling1970} to 1.4, 0.6 and \SI{2.7}{\percent} on 80, 40
and 20 cells,
whichever isotope is carried as the first fluid. The friction
$-\rho_a\nu_{ab}(\vec u_a-\vec u_b)$ between the two fluids uses the
Chapman--Enskog momentum-transfer frequency
$\nu_{ab}=\tfrac43\,n_b\langle\sigma v\rangle\,m_b/(m_a+m_b)$, which makes $\rho_a\nu_{ab}$ symmetric
in the pair and gives the interdiffusion coefficient $D_\mathrm{CE}$.

\emph{Thermal-diffusion equilibrium.} The same mixture at rest in a closed
cylinder at \SI{0.2}{Torr} has an imposed temperature
$T=\SI{300}{\kelvin}+(T_\mathrm{axis}-\SI{300}{\kelvin})(1-r^{2}/R^{2})$.
The composition relaxes to the equilibrium of the thermal force,
$\mathrm{d}\ln(x_\mathrm{D}/x_\mathrm{H})=-\alpha_T(T)\,\mathrm{d}\ln T$, with
the thermal-diffusion factor $\alpha_T(T)$ of the model, which changes sign
near \SI{6000}{\kelvin}. The solver reproduces it to $4\times10^{-4}$ in
$\ln(x_\mathrm{D}/x_\mathrm{H})$ for $T_\mathrm{axis}=\SI{3000}{\kelvin}$
and to $4\times10^{-4}$ for $T_\mathrm{axis}=\SI{12000}{\kelvin}$, which
crosses the sign change; these are 0.13 and \SI{0.11}{\percent} of the range of the
equilibrium profile.

\emph{Two differences from the intended operators.} First, the wall slip is not
Maxwell's condition. On a curved wall that condition contains the shear rate,
$u_\mathrm{s}=-L_\mathrm{s}(\partial_ru_\theta-u_\theta/r)$~\cite{SunBarberEmerson2005},
and the solver omits the term $u_\theta/r$. At a Knudsen number of 0.07 the
slip velocity is then \SI{4}{\percent} ($\sigma_v=1$) and \SI{8}{\percent}
($\sigma_v=0.5$) too small; the difference is negligible below a Knudsen
number of 0.01. Second, heat conduction is solved implicitly and the viscous
heat is deposited afterward. A cell cooled by a wall therefore ends each
step with its temperature excess multiplied by $1+g\,\Delta t/(\rho c_v)$,
with $g$ the conductance to the wall per unit volume. This is an error of
first order in the time step that affects only the row of cells next to a
wall. In a test with wall loss the excess is \SI{8}{\percent} at
$\Delta t=\SI{1}{\micro\second}$, of which \SI{4}{\percent} is this factor
and the rest an artifact of the driven core of the test.

\subsection{Spin-down against walls at rest}
\label{app:neutral-tests-spindown}

\begin{figure*}[tp]
\includegraphics[width=\textwidth]{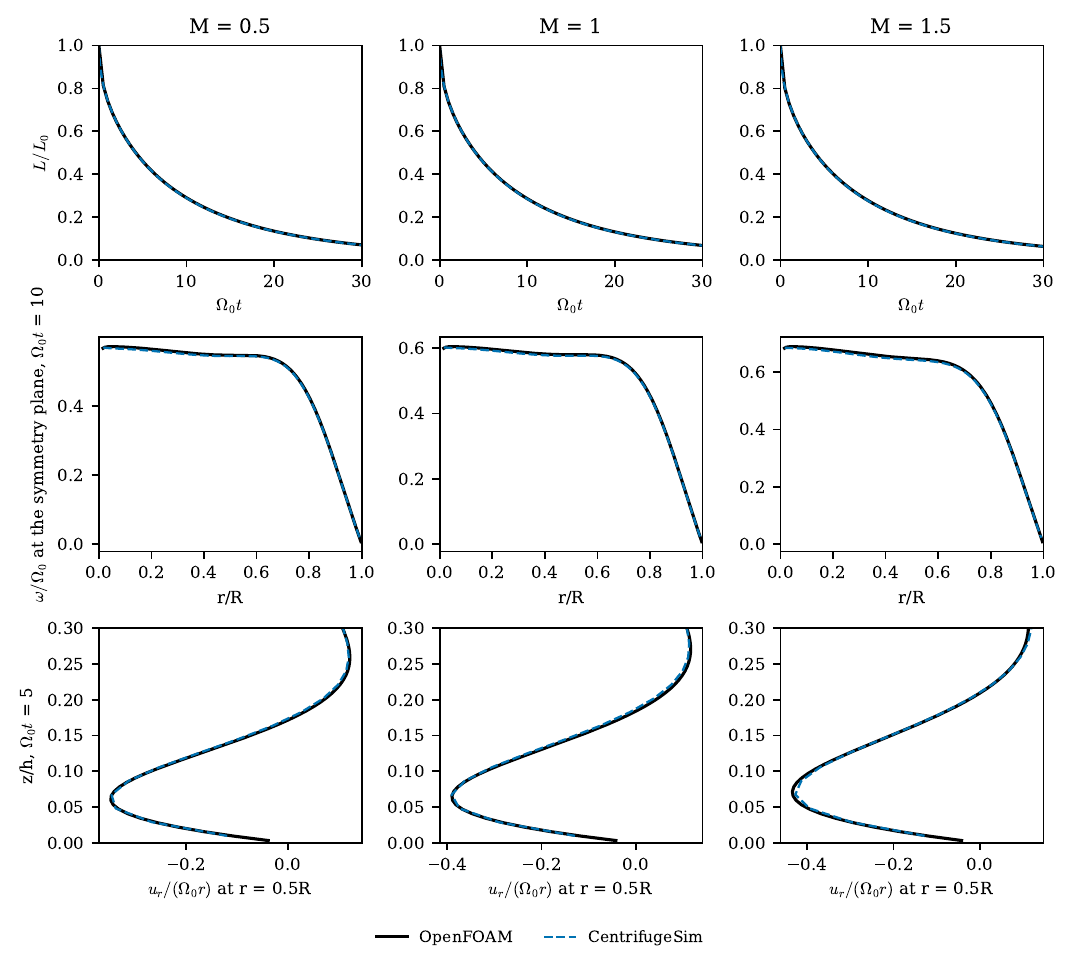}
\caption{\label{fig:neutral-tests-spindown}Spin-down of a rigidly rotating gas
against walls at rest, $Re=400$, $h=R$: the gas solver in the configuration of
the simulations (dashed) against OpenFOAM (solid). Top: angular momentum of the
gas. Middle: angular velocity at the symmetry plane at $\Omega_0t=10$. Bottom:
radial velocity in the plate layer at $r=0.5R$ and $\Omega_0t=5$.}
\end{figure*}

A gas in rigid rotation $\Omega_0$, in isothermal centrifugal equilibrium at
\SI{300}{\kelvin}, is left to spin down against the plate and the side wall,
both at rest from $t=0$ ($Re=400$, $h=R$, $\delta=R/20$). Nothing drives the
flow: the layer on the plate forms, pumps gas inward, and the circulation
carries the angular momentum to the walls faster than viscous diffusion
alone. At $M\geq1$ the gas is compressed toward the wall and heated by
dissipation. The comparison is made at $M=0.5$, 1 and 1.5. OpenFOAM is run on
$40^2$, $80^2$ and $160^2$ uniform cells; its finest grid is within
\SI{0.2}{\percent} of the extrapolated result for the angular momentum, the
core rotation and the temperature and within \SI{0.7}{\percent} for the inflow
peak, and halving its time step changes them by less than \SI{0.2}{\percent}.
The measures are the time $T_{1/4}$ at which the angular momentum of the gas
has fallen to a quarter, the angular velocity of the core, the peak of the
radial inflow in the plate layer at $r=0.5R$ and $\Omega_0t=5$, and the largest
gas temperature (Fig.~\ref{fig:neutral-tests-spindown},
Table~\ref{tab:neutral-tests-spindown}). The legends of
Figs.~\ref{fig:neutral-tests-spindown}--\ref{fig:neutral-tests-hotcore} call the gas solver CentrifugeSim.

\begin{table}[b]
\centering
\caption{Spin-down against walls at rest: OpenFOAM, and the difference of the
gas solver from it. Core rotation: angular velocity at $0.3R$ near the
symmetry plane at $\Omega_0t=10$.}
\label{tab:neutral-tests-spindown}
\setlength{\tabcolsep}{3pt}\footnotesize
\begin{tabular}{lccc}
\hline\hline
$M$ & 0.5 & 1 & 1.5 \\
\hline
\multicolumn{4}{l}{OpenFOAM} \\
\quad $\Omega_0T_{1/4}$ & 11.78 & 11.58 & 11.18 \\
\quad peak inflow ($\Omega_0r$) & 0.349 & 0.391 & 0.435 \\
\quad peak temperature (K) & 311 & 348 & 411 \\
\multicolumn{4}{l}{Gas solver, difference (\%)} \\
\quad $T_{1/4}$ & $-0.6$ & $-0.6$ & $-0.7$ \\
\quad peak inflow & $-1.2$ & $-0.8$ & $-2.4$ \\
\quad core rotation & $-0.7$ & $-0.7$ & $-0.9$ \\
\quad peak temperature & $0.0$ & $+0.1$ & $+0.2$ \\
\hline\hline
\end{tabular}
\end{table}

The angular momentum of the gas follows the OpenFOAM result throughout: the
spin-down time is 0.6--\SI{0.7}{\percent} short, the core rotation within
\SI{0.9}{\percent} and the peak temperature within \SI{0.2}{\percent}. The
inflow of the plate layer is 0.8--\SI{2.4}{\percent} weak, the most at
$M=1.5$.

\subsection{End-plate layer under a rotating core}
\label{app:neutral-tests-layer}

\begin{figure*}[tp]
\includegraphics[width=\textwidth]{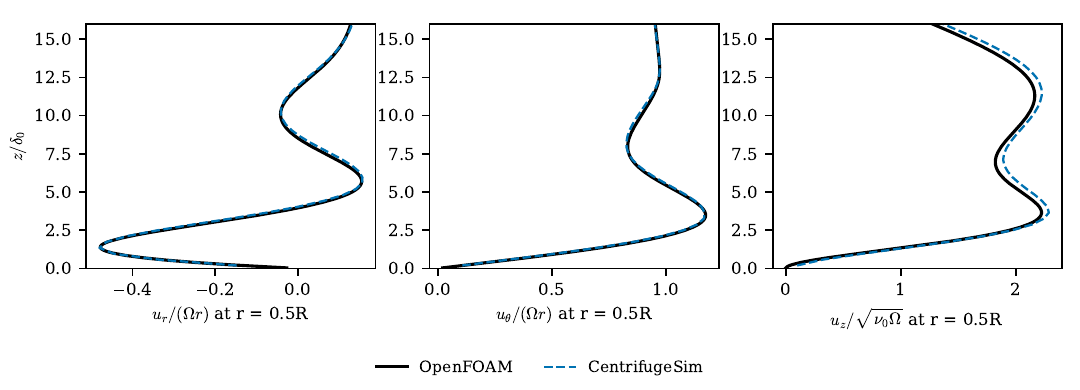}
\caption{\label{fig:neutral-tests-layer}End-plate layer under a core held in
rigid rotation above $12\,\delta_0$ ($M=1.3$, $Re=6400$), at $\Omega t=30$ and
$r=0.5R$: radial, azimuthal and axial velocity against height. Gas solver in
the configuration of the simulations (dashed) against OpenFOAM (solid).}
\end{figure*}

To obtain a steady layer, the gas above $12\,\delta_0$ ($\delta_0=\sqrt{\nu_0/\Omega}$, with $\nu_0$ the kinematic viscosity at the wall temperature and the fill density $\rho_0$) is held in rigid
rotation by an azimuthal force $\rho k(z)(\Omega r-u_\theta)$, with
$k=5\Omega$ above that height and a $\tanh$ transition of width $\delta_0$;
below it only the walls act ($h=20\,\delta_0$, symmetry plane at $z=h$, walls
at \SI{300}{\kelvin}). The force and its work enter both codes identically.
The case is built at the numbers of the end-plate layer in the simulations,
$M=1.3$ and $Re=6400$ ($\delta_0=R/80$): there the layer lies in the cooler,
denser gas next to the plate, where $Re=6\times10^{3}$--$1.6\times10^{4}$.
OpenFOAM is run on three graded grids, the finest of which
changes the measures below by less than \SI{0.5}{\percent} from the next. The
flow is compared at $\Omega t=30$, when it is steady, in the variables of the
layer under a rotating fluid~\cite{Bodewadt1940}: $u_r/(\Omega r)$ and
$u_\theta/(\Omega r)$ against $z/\delta_0$, and the pumping, the largest
upward mass flux through the disk of radius $r$ below $10\,\delta_0$ divided
by $\pi r^{2}\rho_0\sqrt{\nu_0\Omega}$
(Fig.~\ref{fig:neutral-tests-layer}, Table~\ref{tab:neutral-tests-layer}).

\begin{table}[b]
\centering
\caption{End-plate layer under a rotating core
(Section~\ref{app:neutral-tests-layer}) and under a hot core
(Section~\ref{app:neutral-tests-hotcore}): OpenFOAM, and the difference of the
gas solver from it. Inflow: magnitude of the peak radial inflow (positive:
stronger); $u_\theta$:
peak azimuthal velocity; both at $0.5R$. Pumping: see the text.}
\label{tab:neutral-tests-layer}
\setlength{\tabcolsep}{2pt}\footnotesize
\begin{tabular}{lcccc}
\hline\hline
 & \multicolumn{2}{c}{Rotating core} & \multicolumn{2}{c}{Hot core} \\
 & OpenFOAM & diff. & OpenFOAM & diff. \\
\hline
Inflow ($\Omega r$) & 0.477 & $+0.5\,\%$ & 0.160 & $+3.9\,\%$ \\
$u_\theta$ ($\Omega r$) & 1.175 & $-0.4\,\%$ & 0.918 & $-3.0\,\%$ \\
Pumping $0.3R$ & 1.83 & $+6.3\,\%$ & 2.23 & $+2.4\,\%$ \\
Pumping $0.5R$ & 1.95 & $+4.6\,\%$ & 1.90 & $+3.4\,\%$ \\
Pumping $0.7R$ & 1.69 & $+2.9\,\%$ & 1.71 & $+3.1\,\%$ \\
Temperature (K) & 417 & $+1.2\,\%$ & 11\,989 & $-0.2\,\%$ \\
\hline\hline
\end{tabular}
\end{table}

The layer is reproduced: the inflow and the azimuthal velocity are within
\SI{1}{\percent} and the pumping within 2.9--\SI{6.3}{\percent}, the largest at
$0.3R$; above the layer (Fig.~\ref{fig:neutral-tests-layer}) the second
inflow near $10\,\delta_0$ is slightly weaker than in OpenFOAM. The peak
temperature is \SI{1.2}{\percent} high.

\subsection{End-plate layer under a hot core}
\label{app:neutral-tests-hotcore}

\begin{figure*}[tp]
\includegraphics[width=\textwidth]{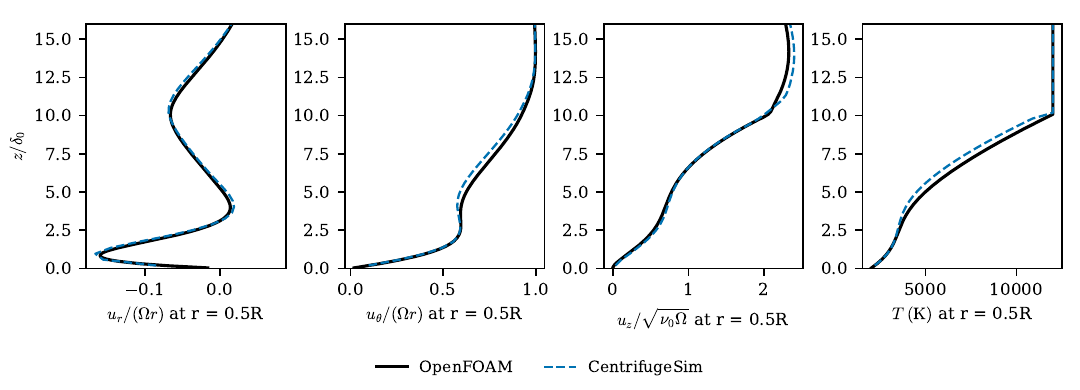}
\caption{\label{fig:neutral-tests-hotcore}End-plate layer under a hot core
(walls at \SI{2000}{\kelvin}, gas above $10\,\delta_0$ held at
\SI{12000}{\kelvin}, rotation held above $12\,\delta_0$; $M=1.3$, $Re=6400$
at the wall temperature) at $\Omega t=30$ and $r=0.5R$: velocities and
temperature against height. Gas solver in the configuration of the
simulations (dashed) against OpenFOAM (solid).}
\end{figure*}

In the simulations the layer lies under a hot, viscous core. The previous
case is repeated with the walls at \SI{2000}{\kelvin} ($M=1.3$ and $Re=6400$
at that temperature, $h=40\,\delta_0$) and the temperature held above
$10\,\delta_0$, a cell face in both codes. Since rigid rotation with a jump
of temperature is not an equilibrium, the held temperature is raised from
\SI{2000}{\kelvin} to \SI{12000}{\kelvin} over $\Omega t=0$--5. The kinematic
viscosity of the core is then about 20 times that of the layer, the gas in the
layer is heated by conduction from above as well as by dissipation, and the
meridional flow is slow, $u_r\approx0.16\,\Omega r$ at its peak, so that the
numerical dissipation of the flux matters more than in the previous case.
OpenFOAM's finest grid is within \SI{1}{\percent} of the extrapolated values.
The pumping is within 2.4--\SI{3.4}{\percent}, the azimuthal velocity within
\SI{3.0}{\percent}, the inflow \SI{3.9}{\percent} strong and the temperature
within \SI{0.3}{\percent}
(Fig.~\ref{fig:neutral-tests-hotcore}, Table~\ref{tab:neutral-tests-layer}).

\section*{References}
\bibliography{references}